\documentstyle[11pt,aaspp4]{article}
\lefthead{Rand}
\righthead{Spectroscopy of DIG in NGC 891}

\slugcomment{}

\begin{document}

\title{Further Spectroscopy of the Diffuse Ionized Gas in NGC 891 and
Evidence for a Secondary Source of Ionization}

\author{Richard J. Rand\altaffilmark{1}}
\affil{Univ. of New Mexico, Dept. of Physics and
Astronomy, 800 Yale Blvd, NE, Albuquerque, NM 87131}
\authoremail{rjr@gromit.phys.unm.edu}
\altaffiltext{1}{Visiting Astronomer, National Optical Astronomy Observatories,
Tucson, AZ}

\begin{abstract}

Two long-slit spectra of the diffuse ionized gas in NGC 891 are
presented.  The first reveals variations parallel to the major axis in
emission line ratios in the halo gas at $z=700$ pc.  It is found that
filaments of H$\alpha$ emission show lower values of
[N$\,$II]/H$\alpha$, [S$\,$II]/H$\alpha$ and [O$\,$I]/H$\alpha$.
Although this result is expected if the filaments represent the walls
of evacuated chimneys, it merely reflects a more general correlation
of these ratios with H$\alpha$ surface brightness along the slit, and
may simply arise from radiation dilution effects.  Halo regions
showing low line ratios are probably relatively close to ionizing
sources in the disk below.  The results highlight difficulties
inherent in observations of edge-on galaxies caused by lack of
knowledge of structure in the in-plane directions.  The
[S$\,$II]/[N$\,$II] ratio shows almost no dependence on distance along
the major axis or H$\alpha$ surface brightness.  Values of
[O$\,$I]/H$\alpha$ indicate that H is 80--95\% ionized
(assuming $T=10^4$ K), with the higher ionization fractions correlating
with higher surface brightness.

Much more interesting information on the nature of this gaseous halo
comes from the second observation, which shows the vertical dependence
of [N$\,$II]/H$\alpha$, [S$\,$II]/H$\alpha$, [O$\,$I]/H$\alpha$, and
[O$\,$III]/H$\beta$ through the brightest region of the DIG halo.  The
most surprising result, in complete contradiction to models in which
the DIG is ionized by massive stars in the disk, is that
[O$\,$III]/H$\beta$ rises with height above the plane for $z>1$ kpc
(even as [N$\,$II]/H$\alpha$, [S$\,$II]/H$\alpha$, and
[O$\,$I]/H$\alpha$ are rising, in line with expectations from such
models).  The run of [S$\,$II]/[N$\,$II] is also problematic, showing
essentially no contrast with $z$.  The [O$\,$III] emission probably
arises from shocks, turbulent mixing layers, or some other secondary
source of ionization.  Composite models in which the line emission
comes from a mix of photo-ionized gas and shocks or turbulent mixing
layers are considered in diagnostic diagrams, with the result that
many aspects of the data can be explained.  Problems with the run of
[S$\,$II]/[N$\,$II] still remain, however.  There is a reasonably
large parameter space allowed for the second component.  For the
photo-ionized component, only matter-bounded models succeed, putting a
fairly strong restriction on the clumpiness of the halo gas.  Given
the many uncertainties, the composite models can do little more than
demonstrate the feasibility of these processes as secondary sources of
energy input.  A fairly robust result, however, is that the fraction
of H$\alpha$ emission arising from the second component probably
increases with $z$.  From values of [O$\,$I]/H$\alpha$, H is
essentially 100\% ionized at $z=0$ kpc and 90\% ionized at $z=1$ kpc
(again assuming $T=10^4$ K).

\end{abstract}


\section{Introduction}

The vast majority of the free electrons in the ISM of the Milky Way
reside in a thick ($\sim$ 900-pc scale height) diffuse layer known as
the Reynolds layer or the Warm Ionized Medium (e.g. Reynolds 1993).
This phase fills about 20\% of the ISM volume, with a local midplane
density of about 0.1 cm$^{-3}$.  Such a phase is now known to be a
general feature of external star-forming galaxies, both spirals
(e.g. Walterbos 1997; Rand 1996) and irregulars (e.g. Hunter \&
Gallagher 1990; Martin 1997, hereafter M97), where it is commonly
referred to as Diffuse Ionized Gas (DIG).  However, for edge-on
spirals, only in the more actively star-forming galaxies does the gas
manifest itself as a smooth, widespread layer of emission detectable
{\it above} the HII region layer (Rand 1996).  One such galaxy, NGC
891, is an attractive target for study, not only because of its
prominent DIG layer (Rand, Kulkarni, \& Hester 1990; Dettmar 1990),
but also its proximity ($D=9.5$ Mpc will be assumed here) and nearly
fully edge-on aspect ($i>88\arcdeg$; Swaters 1994).

One of the outstanding problems in the astrophysics of the ISM is the
ionization of these layers.  For the Reynolds layer, the local
ionization requirement ($5\times 10^6$ s$^{-1}$ per cm$^{2}$ of
Galactic disk; Reynolds 1992) is comfortably exceeded (by a factor of
6 or 7) only by the ionizing output of massive stars.  Alternatively,
the ionization would require essentially all the power put out by
supernovae (Reynolds 1984) -- hence, this energy source could
contribute at some level but probably cannot explain all of the
diffuse emission.  Photo-ionization models, on the other hand, must
explain how the ionizing photons can travel $\sim$ 1 kpc or more from
their origin in the thin disk of massive stars to maintain this
distended layer.

Crucial information on both the ionization and thermal balance of DIG
comes from emission line ratios.  In the Reynolds layer, ratios of
[S$\,$II] $\lambda\lambda6716,6731$ and [N$\,$II]
$\lambda\lambda6548,6583$ to H$\alpha$ are generally enhanced relative
to their HII-region values, while [O$\,$III] $\lambda5007$/H$\alpha$
is much weaker.  These contrasts are in accordance with models in
which photons leak out of HII regions and ionize a larger volume, with
the radiation field becoming increasingly diluted with distance from
the HII region [Mathis 1986; Domg$\ddot {\rm o}$rgen, \& Mathis 1994;
Sokolowski 1994 (hereafter S94; see also Bland-Hawthorn, Freeman, \&
Quinn 1997)].  The effect of this dilution, measured by the ionization
parameter, $U$, is primarily to allow species such as S and O, which
are predominantly doubly ionized in HII regions, to recombine into a
singly ionized state.  The effect may be less noticeable for N
because it is mostly singly ionized in HII regions.  The Wisconsin
H$\alpha$ Mapper (WHAM) has been used to determine [O$\,$I]/H$\alpha$
in three low-latitude directions, resulting in values $<0.01$ to 0.04
(Haffner \& Reynolds 1997).  Such low values imply, since the
ionization of O and H are strongly coupled by a charge
exchange reaction, that the diffuse gas is nearly completely ionized
(Reynolds 1989).

Although weak, [O$\,$III] emission has been detected in two directions
in the Reynolds layer at $b=0\arcdeg$ (Reynolds 1985), with the result
[O$\,$III]/H$\alpha = 0.06$.  Reynolds postulated that the [O$\,$III]
emission does not arise from diluted stellar ionization but from gas
at about 10$^5$ K, presumably the same gas as seen in C$\,$ IV
$\lambda$1550 and O$\,$III] $\lambda1663$ emission by Martin \& Bowyer
(1990).  The origin of this rapidly cooling gas is unclear.
[O$\,$III] emission from the DIG of NGC 891 and the implications for
DIG ionization is one of the main subjects of this paper.

In external spiral galaxies, smooth increases in [S$\,$II]/H$\alpha$
and [N$\,$II]/H$\alpha$ vs. distance from HII regions have been
observed in both the in-plane and vertical directions in accordance
with photo-ionization models [Walterbos \& Braun 1994; Dettmar \&
Schulz 1992; Rand 1997a (hereafter R97); Golla, Dettmar, \& Domg$\ddot
{\rm o}$rgen 1996; Greenawalt, Walterbos, \& Braun 1997; Wang,
Heckman, \& Lehnert 1997].  The same trends are seen in irregulars
(Hunter \& Gallagher 1990; M97).  This behavior has been revealed in
NGC 891 through spectra using long slits oriented vertically to the
plane.  Dettmar \& Schulz (1992) placed a slit at $R=65''$ NE of the
nucleus, while R97 took a deeper spectrum at $R=100''$ NE.  R97 found
that [N$\,$II]/H$\alpha$ rises to a value of 1.4, implying a very hard
ionizing spectrum.  S94, which pays particular attention to modeling the
DIG of NGC 891, can predict such a high value only by assuming
a stellar IMF extending to 120 M$_{\sun}$, a reduction in cooling
efficiency due to elemental depletions, and hardening of the radiation
field by the intervening gas.  [O$\,$I]/H$\alpha$ was not detected by
Dettmar \& Schulz (1992) in the halo of NGC 891 at $R=65''$ NE, with
an upper limit of 0.05.  Dettmar (1992) also reported an upper limit
on [O$\,$III]/H$\beta$ of 0.4 at the same location.

A wealth of forbidden-line long-slit data on bright DIG and HII
regions in irregular galaxies has recently been published by M97.
Through the use of line-diagnostic diagrams (e.g.  Baldwin, Phillips,
\& Terlevich 1981; Veilleux \& Osterbrock 1987), she finds that while
photo-ionization models can explain the line ratio behavior in many
galaxies, the rather shallow fall-off of [O$\,$III]/H$\beta$ with
distance from HII regions and the sharp rise in [O$\,$I]/H$\alpha$
seen in some galaxies imply a second source of ionization.  Shocks are
favored as the most likely second source.

The forbidden lines, though bright, are sensitive to metallicity and
temperature and thus their interpretation in terms of ionization
scenarios is complicated by uncertainties in abundances, degree of
depletion, and sources of non-ionization heating.  A more direct
constraint on the ionizing spectrum has come from the very weak
He$\,$I $\lambda$5876 line.  He$\,$I/H$\alpha$ is relatively easy to
interpret in terms of the ratio of helium- to hydrogen-ionizing
photons, allowing the hardness of the ionizing spectrum, the mean
spectral type of the responsible stars, and the upper IMF cutoff to be
inferred, assuming pure stellar photoionization.  The results for the
Reynolds layer (Reynolds \& Tufte 1995), for HI worms from equivalent
radio recombination lines (Heiles et al. 1996) and for NGC 891 (R97)
all imply a much softer spectrum than do the forbidden lines.  Further
consequences of this discrepancy are discussed in the above three
references.

The goal of this paper is to make further progress in understanding
the ionization of DIG in spirals.  The motivations are two-fold.
First, the DIG halo of NGC 891 features several bright filaments and
shells.  It is likely that some of these are chimney walls (Norman \&
Ikeuchi 1989) surrounding regions of space evacuated by many
supernovae.  In this case, radiation from any continuing star
formation near the base of the chimney will have an unimpeded journey
to the walls, and thus the filaments may be directly ionized and show
a spectrum more like an HII region than diffuse gas, which receives a
significant contribution from relatively soft diffuse re-radiation
(Norman 1991).  If true, then the filaments should show lower
[N$\,$II]/H$\alpha$ and [S$\,$II]/H$\alpha$ than the surrounding gas.
To this end, a spectrum has been taken with a slit oriented parallel
to the major axis, but offset into the halo gas, traversing several
filaments.

The second purpose is to study in more detail the dependence of line
ratios on $z$ beyond the results reported in R97 for
He$\,$I/H$\alpha$ and [N$\,$II]/H$\alpha$.  By adding measurements of
[S$\,$II], [O$\,$III], [O$\,$I], and H$\beta$, one can form diagnostic
diagrams and thus constrain the source(s) of ionization in the spirit
of M97.

\section{Observations}

The spectra were obtained at the KPNO 4-m telescope on 1996 December
12--13.  The slit positions are shown in Figure 1 (Plate 00) overlaid
on the H$\alpha$ image of \markcite{}RKH and on a version of the image
in which a median filter has been applied and the resulting smooth
image subtracted to reveal the filaments clearly.  Note also how the
filaments connect onto the brightest HII regions in the disk, with few
exceptions.  For the first night, the slit was oriented parallel to
the major axis and offset from it by 15'' along the SE side of the
minor axis.  This slit position will be referred to as the \lq\lq
parallel slit.\rq\rq\ \ The slit position for the second night was the
same as in R97's observations: oriented perpendicular to the major
axis and centered $R=100"$ on the NW (approaching) side of NGC 891 --
the \lq\lq perpendicular slit.\rq\rq\ \ The slit length is 5', and the
spatial scale is 0.69" per pixel.

The KPC-24 grating was used with the T2KB 2048x2048 CCD, providing a
dispersion of 0.53$\AA$ per pixel, a resolution of 1.3$\AA$, and a
useful coverage of about 800 $\AA$.  For observations of red lines,
the grating was tilted to give a central wavelength of about
6600$\AA$, allowing H$\alpha$, [N$\,$II] $\lambda\lambda6548,6583$,
[S$\,$II] $\lambda\lambda6716,6731$, and [O$\,$I] $\lambda6300$ to be
observed.  For the blue lines [O$\,$III] $\lambda 5007$ and H$\beta$,
the central wavelength was set to about 5000$\AA$.

For the parallel slit, seven half-hour spectra were taken, along with
separate sky exposures because no part of the slit covered pure sky.
For the perpendicular slit, seven half-hour spectra were taken
covering the blue lines, and three covering the red lines.  Sky
subtraction for these spectra was achieved using regions of pure sky
at both ends of the slit.  Exact slit center positions were varied to
allow removal of chip defects in the stacking process.

The reduction was carried out with the IRAF\footnote{IRAF is
distributed by the National Optical Astronomy Observatories, which is
operated by the Association for Research in Astronomy, Inc., under
cooperative agreement with the National Science Foundation.} package.
Small-scale variations in response were removed using projector flats.
The slit illumination correction was determined with sky flats, and
the spectral response function with standard stars.  Arc lamp
exposures were used to calibrate the wavelength scale as a function of
location along the slit.  The final calibrated, sky-subtracted spectra
were spatially aligned and stacked.  The noise in continuum-free
regions of the stacked perpendicular slit exposures is $2\times
10^{-18}$ and $3\times 10^{-18}$ erg cm$^{-2}$ s$^{-1}$ $\AA^{-1}$
arcsec$^{-2}$ in the blue and red spectra, respectively.  For the
stacked parallel slit exposures, continuum covers the entire slit
length and consequently the pixel-to-pixel variations are higher:
$6\times 10^{-18}$ erg cm$^{-2}$ s$^{-1}$ $\AA^{-1}$ arcsec$^{-2}$.
Since the spectral lines appear well represented by Gaussians, line
properties were determined with Gaussian fits.  The continuum level
was estimated from a linear fit to the continuum on each side of the
line.  Error bars reflect the noise in the spectra and the uncertainty
in the fit of the spectral response function.

Line ratios for the perpendicular slit are consistent with those in
R97, given the calibration and slit positioning uncertainties.
However, there is a mistake in the velocity scale of Figure 10 of R97:
all velocities are about 35 km s$^{-1}$ too low.

\section{Results and Discussion}

\subsection{The \lq\lq Parallel\rq\rq\ Slit}

The H$\alpha$, [N$\,$II], and [S$\,$II] lines were detected along the
entire useable slit length, which corresponds to about 15 kpc at the
assumed distance of NGC 891.  Reliable parameters could be measured
over a 13 kpc region.  [O$\,$I] is detected from about 3 kpc SW of the
peak in continuum emission to the NE edge of the slit.  However,
confusion with a sky line limits measurement of reliable parameters on
the NE side to a maximum distance of about 5 kpc from the continuum
peak.  Figure 2 shows the runs of [S$\,$II] $\lambda 6716$/H$\alpha$,
[N$\,$II] $\lambda$6583/H$\alpha$, [O$\,$I] $\lambda 6300$/H$\alpha$,
[S$\,$II] $\lambda 6716$/[N$\,$II]$\lambda$6583, and normalized
H$\alpha$ surface brightness with position along the slit.  The data
have been averaged over 10 pixels, or about 300 pc, except in the case
of [O$\,$I]/H$\alpha$, where the averaging is over 20 pixels.  Figure
1b shows that the slit traverses four bright filaments.  These are
apparent as peaks in the H$\alpha$ profile (marked in Figure 2a).  The
emission is also obviously much brighter on the NE side than on the SW
side, confirming the result from the images of RKH and Dettmar (1990).
The ratio of the two [S$\,$II] lines is consistent with the
low-density limit of 1.5 (Osterbrock 1989), even for the filaments.

At first glance, it would seem that the data support the expectation
outlined in \S I: there is a definite reduction in [S$\,$II]/H$\alpha$
and [N$\,$II]/H$\alpha$ at the positions of the four bright filaments.
[O$\,$I] $\lambda 6300$/H$\alpha$ also tends to be lower at the
filament crossings.  However, this result reflects a more general
trend of these line ratios with H$\alpha$ surface brightness, as shown
in Figure 3.  The ratios at the positions of the three brightest
filaments define the correlation at H$\alpha$ surface brightnesses
$\gtrsim 20$ erg cm$^{-2}$ s$^{-1}$ arcsec$^{-2}$.  There is no
evidence for a discontinuity or steepening of the correlation at these
intensities.  In fact, the slope becomes nearly flat here.

What is the reason for these very good correlations?  Very similar
results are found for the perpendicular slit data (Figure 7),
suggesting, in a stellar photo-ionization scenario, that they reflect
the well-established variation (e.g. S94) of line ratios with
ionization parameter, $U$, which measures the diluteness of the
radiation field (see \S 1).  If this is the case, then it is implied
that $U$ is lower along lines of sight with faint DIG, even when
comparing at the same $z$.

Gas along lines of sight with faint H$\alpha$ emission may be
relatively remote from ionizing stars in the disk, so that large
columns of intervening gas and geometric dilution result in a low $U$.
Alternatively, it is possible that such gas is no more remote from
ionizing stars, but that $U$ is low because the responsible clusters
feature fewer such stars, leading to an intrinsically weak emergent
ionizing radiation field.  The fact that the lines of sight passing
through the filaments feature the brightest DIG while the filaments
are clearly associated with bright visible HII regions in the disk
(Figure 1b) would tend to suggest that the former explanation is
correct.  However, the dust lane may be hiding numerous fainter HII
regions whose stars may be the primary source of ionization for gas
along lines of sight with fainter emission.  On the other hand, such
HII regions should contain fewer massive stars and have, if the IMF
varies little, a lower probability of containing the most massive
stars, resulting in softer spectra on average.  A softer spectrum
leads to lower line ratios (S94) and would offset the dilution effect
to some degree.  This question will probably be resolved from studies
of DIG in more face-on galaxies where the in-plane variations of
H$\alpha$ surface brightness and line ratios can be related to the
distribution of ionizing stars.

Regardless of the explanation, since the filaments in NGC 891 follow
the overall correlation in Figure 3, it cannot be claimed that the
line ratios are lower in the filaments because they surround {\it
evacuated} regions and are directly ionized by HII regions below.
This does not imply that the filaments are not chimney walls, but does
point out the potential difficulties in deriving information on
isolated structures in edge-on galaxies.

It is interesting that [S$\,$II]/[N$\,$II] shows almost no spatial
variation or dependence on H$\alpha$ surface brightness compared to
the other line ratios.  This ratio will be discussed further in the
next section, where it will become clear that there is little
dependence on $z$ either.

Figure 4 shows heliocentric velocity centroids, formed from a weighted
average of the H$\alpha$, [S$\,$II], and [N$\,$II] line centroids,
along with the emission profile.  One of the filaments on the NE side
and a broad region centered on a filament on the SW side show
velocities further from the systemic velocity than expected from the
smooth, nearly linear trend of velocity with position.  Again, this is
not necessarily an indication that the filaments have peculiar
velocities, but may simply indicate that they are located in the inner
disk.  If so, then these velocities are more heavily weighted by inner disk
material than are adjacent ones.  Inner disk gas will show velocities
further from the systemic velocity compared to outer disk gas because
of the greater projection of the rotation velocity vector along the
line of sight.  Hence, the velocity deviations may simply be due to
geometrical effects.

\subsection{The \lq\lq Perpendicular\rq\rq\ Slit}

The H$\alpha$ line, the [N$\,$II] $\lambda 6583$ line and the
[S$\,$II] lines are detected up to about $z=3$ kpc on each side of the
plane.  [O$\,$I] is detected to about half this height, while
[O$\,$III] and H$\beta$ are detected up to about $z=2.5$ kpc.

Shown in Figure 6 are the vertical runs of [S$\,$II] $\lambda
6716$/H$\alpha$, [N$\,$II] $\lambda 6583$/H$\alpha$,
[O$\,$I]/H$\alpha$, [S$\,$II] $\lambda 6716$/[N$\,$II] $\lambda 6583$,
and [O$\,$III]/H$\beta$.  The H$\alpha$ profile is also plotted -- the
local minimum at $z=0$ kpc is due to the dust lane.  The data are
averaged over 10 pixels.  [N$\,$II]/H$\alpha$ and [S$\,$II]/H$\alpha$
show a smooth and remarkably similar increase with $z$, from about
0.35 in the midplane to over 1.0 at $z=2-3$ kpc.  [S$\,$II]/[N$\,$II]
is nearly constant at about 0.6 at all $z$ where the uncertainties are
not too large.  [O$\,$I]/H$\alpha$ increases from about 0.03 at $z=0$
kpc to 0.08 at $z=1.3$ kpc.  However, the most surprising result is
that [O$\,$III]/H$\beta$ {\it rises} from 0.3 in the midplane to 0.8
at $z=2$ kpc.  Figure 7 shows the correlation of these ratios with
H$\alpha$ surface brightness.  These are very similar to the
correlations in Figure 3.  Again, the ratio of the two [S$\,$II] lines
is everywhere consistent with the low-density limit of 1.5.

Again assuming $T_e=10^4\,$K, [O$\,$I]/H$\alpha$ values imply H
is essentially 100\% ionized at $z=0$ kpc, decreasing to about 90\% at
$z=1$ kpc.  If $T_e=8000\,$K, then these ionization fractions are 90\%
and 80\%.

Except for the discrepancy mentioned in \S 2, the mean velocities of
the lines are consistent with the results of R97 and give no
additional information.  Therefore, we will not discuss the kinematics
further.

\subsubsection{Photo-ionization Models}

We now attempt to understand whether the above emission line
properties can be understood by massive-star photo-ionization alone.
In doing so, we temporarily ignore the problems posed by the low
He$\,$I/H$\alpha$ but will return briefly to the reconciliation of
this ratio with the forbidden line ratios in \S 4.  We use unpublished
models from S94 since they are the only models which specifically
attempt to reproduce the line ratios in the DIG of NGC 891.  In these
models an ionizing spectrum of radiation from a population of stars
with an IMF slope of --2.7 (intermediate between Salpeter and
Miller-Scalo values) and stellar atmospheres from Kurucz (1979) is
considered.  This radiation field is allowed to propagate through a
slab of gas (representing a clump of halo gas) in a one-dimensional
calculation, the dilution being measured by the ionization parameter,
$U$, at the front of the slab.  As discussed in \S 1, as lower values
of $U$ are considered, the predominant ionization state of S and O
(and N to a lesser extent) changes from doubly to singly ionized.  As
a consequence, [N$\,$II]/H$\alpha$ and [S$\,$II]/H$\alpha$ rise while
[O$\,$III]/H$\beta$ falls.  At the end of the model slab, an
increasingly neutral zone appears as lower values of $U$ are
considered, leading to a slow rise in [O$\,$I]/H$\alpha$ with $U$.  In
this one-dimensional calculation of pure photo-ionization, $U$ is
expected to decline exponentially with $z$.  This dependence is
probably more complicated in a real galaxy, although $U$ should
generally fall with increasing height.  Thus, further free parameters
are the value of $U$ at $z=0$ kpc, and the run of $U$ with $z$.

Both radiation bounded models and matter bounded models with various
terminating total atomic hydrogen columns for the clumps were
considered by S94.  We will use only his models with the hardest
stellar spectrum considered (with an upper IMF cutoff of 120
M$_{\sun}$), hardening of the radiation field as it propagates through
the intervening gas before reaching the slab, and heavy element
depletions.  Only these models are able to yield [N$\,$II]/H$\alpha$
and [S$\,$II]/H$\alpha$ in the range 1--1.5.  These models have also
been recently published by Bland-Hawthorn et al. (1997).
We examine as two extremes the matter-bounded model (which will be
referred to as PM) with the lowest terminal hydrogen column considered
for the individual clumps, 2$\times 10^{18}$ cm$^{-2}$, and the
radiation-bounded model (PR).

Figures 8 and 9 show these ratios in diagnostic diagrams of
[O$\,$III]/H$\beta$ and [O$\,$I]/H$\alpha$ vs.
[N$\,$II]/H$\alpha$ and [S$\,$II]/H$\alpha$, and [S$\,$II]/H$\alpha$
vs. [N$\,$II]/H$\alpha$.  Along the sequence of points, $z$
generally increases from 0 kpc at the left end to 2 kpc for plots of
[O$\,$III]/H$\beta$, 1.3 kpc for [O$\,$I]/H$\alpha$, and 3 kpc for
[S$\,$II]/H$\alpha$ vs. [N$\,$II]/H$\alpha$ at the right end.
Models PR and PM are shown in Figures 8 and 9 as the
small open circles joined by solid lines.  Values of log$\,U$ are
marked as explained in the captions.

It is immediately obvious that neither model is a good match to the
data.  Most importantly, the flatness of [O$\,$III]/H$\beta$ below
$z=1$ kpc and its rise with [N$\,$II]/H$\alpha$, [S$\,$II]/H$\alpha$
(and $z$) above $z=1$ kpc is at complete odds with the models.  Also,
the typical value of [O$\,$III]/H$\beta$ is poorly predicted by a
model chosen to match the observed [N$\,$II]/H$\alpha$ and
[S$\,$II]/H$\alpha$.  In model PR, while [N$\,$II]/H$\alpha$ and
[S$\,$II]/H$\alpha$ require log$\,U$ in the range --2.3 to --2.7 at
$z=0$ kpc, and --3.3 to --3.7 at $z=2$ kpc, the predicted
[O$\,$III]/H$\beta$ is $\gtrsim 10$ times that observed for most of
this range of $U$.

In model PM, on the other hand, if we use [N$\,$II]/H$\alpha$ and
[S$\,$II]/H$\alpha$ to set $U$ at $z=0$ kpc, we require log$\,U$ in
the range --3.7 to --4.1, and an extra source of [O$\,$III] emission
at lower $U$.  While such a source can be identified and tested (see
below), it is not clear that such a low value of $U$ should apply to
the midplane, given that the disk contains both HII regions and
diffuse gas.  However, at low $z$ the kinematics indicate that we
receive emission preferentially from the outer disk because of the
absorbing dust layer (R97).  This outer disk gas may, like the
high-$z$ gas, see a relatively dilute radiation field because star
formation is concentrated in the inner disk (Rand 1997b).  In that case the
appropriate $U$ for $z=0$ may be quite low and the gradient of $U$
with $z$ rather shallower than expected in a galaxy without such a
dust lane.

Both models are more successful at reproducing the runs of
[O$\,$I]/H$\alpha$ vs.  [N$\,$II]/H$\alpha$, [O$\,$I]/H$\alpha$
vs. [S$\,$II]/H$\alpha$ and [S$\,$II]/[N$\,$II].  If
[N$\,$II]/H$\alpha$ were lower by about --0.2 dex at low $z$, PR would
fit these data quite well, with log$\,U=-2.6$ at $z=0$ kpc, $-3.3$ at
$z=1$ kpc, and $-3.7$ at $z=2$ kpc.  Model PM would fit equally well
for log$\,U=-4.0$ at $z=0$, $-4.5$ at $z=1$ kpc, and $-5.0$ at $z=2$
kpc.  A similar discrepancy in the observed vs. modeled vertical run
of [S$\,$II]/[N$\,$II] was noted by Golla et al. (1996) in NGC 4631.
[N$\,$II]/H$\alpha$ is expected to show less disk-halo contrast than
[S$\,$II]/H$\alpha$ because the change in predominant ionization state
between HII regions and diffuse gas is smaller for N due to its higher
ionization potential.  This trend is reflected in the models of both
S94 and Domg$\ddot {\rm o}$rgen, \& Mathis (1994).  In both NGC 891
and NGC 4631, however, the disk-halo contrast in [S$\,$II]/H$\alpha$
is rather similar to that in [N$\,$II]/H$\alpha$.

One important factor in explaining the common behavior of
[S$\,$II]/H$\alpha$ and [N$\,$II]/H$\alpha$ might be a {\it radial}
abundance gradient in league with the dust absorption effect noted
above.  Rubin, Ford, \& Whitmore (1984) found that log
([S$\,$II]/[N$\,$II]) generally increases in HII regions in spirals by
0.3 \lq\lq from inner to outer HII regions\rq\rq .  If such a gradient
is present in NGC 891, then the fact that we preferentially observe
outer disk gas at low-$z$ means that [S$\,$II]/[N$\,$II] should be
higher there for a given $U$.  The gradual inclusion of more inner
disk gas with increasing $z$ will then tend to offset the dependence
of [S$\,$II]/[N$\,$II] on $U$ in the models.  However, the parallel
slit data suggests little dependence of the ratio on distance along
the major axis, although there is a slight trend in the right
direction between 2 and 7 kpc from the center on both sides.  Hence,
inasmuch as the major-axis dependence of [S$\,$II]/[N$\,$II] at
$z=700$ pc reflects the radial dependence at $z=0$ pc, it would seem
that an abundance gradient does not affect the line ratios
significantly.  Other expected effects of an abundance gradient are
also not seen in the parallel slit data.  Domg$\ddot {\rm o}$rgen, \&
Mathis (1994) find that [N$\,$II]/H$\alpha$ and [S$\,$II]/H$\alpha$
increase with abundance, while [O$\,$I]/H$\alpha$ shows little
dependence.  In the presence of a radial gradient, these trends, if
strong enough, might be revealed as a decline in [N$\,$II]/H$\alpha$
and [S$\,$II]/H$\alpha$ with distance along the major axis.  The
dominant trend there is with H$\alpha$ brightness and thus it seems
that variations in $U$ are the more important effect.  Again, though,
it must be noted that the averaging along the line of sight in the
parallel slit data will diminish the observable effects of an
abundance gradient.  So far, it has been difficult to find much
variation in [S$\,$II]/[N$\,$II] in the DIG of edge-on galaxies.  This
question remains open to further exploration.

\subsubsection{Composite Photo-ionization and Shock Models}

The most surprising result for the perpendicular slit is the behavior
of [O$\,$III]/H$\beta$ with $z$.  Although such high (and higher)
values of [O$\,$III]/H$\beta$ are found in HII regions, they are a
feature of high-excitation (high $U$) conditions.  If the level of
excitation were increasing with $z$, however, we would also expect to
see [N$\,$II]/H$\alpha$ and [S$\,$II]/H$\alpha$ falling, contrary to
what is observed.  As discussed above, their rise is qualitatively
consistent with a smooth transition from predominantly doubly-ionized
to singly-ionized states, as expected in the dilute photo-ionization
models.  Hence, it is very unlikely that the [O$\,$III] emission
arises from the same DIG component as the [N$\,$II] and [S$\,$II].  We
therefore require a second source of diffuse H$\alpha$ emission which
features bright [0$\,$III].  Two plausible mechanisms for producing
such emission, shocks and turbulent mixing layers are discussed in
this and the next subsection. respectively.  Both can produce bright
[O$\,$III] emission.  These mechanisms were also considered for DIG in
irregular galaxies by M97.  It should be noted, though, that the DIG
in these irregulars is generally much brighter than that studied here.

We first consider whether the line ratio data can be
explained if some of the DIG emission is produced by shock ionization.
We consider only the PM model further because the PR model would
require a second source of H$\alpha$ emission with highly unusual
properties: if the value of $U$ at the midplane is to be roughly
--2.3 to --2.7, then the second component must dominate the
stellar-ionized component by a factor of 30 or more and have
essentially no [O$\,$III] emission if the runs of
[O$\,$III]/H$\beta$ vs. [S$\,$II]/H$\alpha$ and [N$\,$II]/H$\alpha$
are to be explained.  Low-speed shocks do have this property, but also
produce large amounts of [O$\,$I] emission, further confouding the
problem.  Alternatively, if log$\,U\sim -4.0$ at $z=0$ kpc, then the
second component must have essentially no [N$\,$II] or [S$\,$II]
emission and account for about 75\% of the H$\alpha$ emission at $z=0$
kpc in order to reproduce the values of [S$\,$II]/H$\alpha$ and
[N$\,$II]/H$\alpha$.  In either case, it is also difficult to see how
the subsequent rise in all the line ratios with $z$ would be achieved
without the model being highly contrived.  With the PM model, there is
some hope of matching the data by adding a strong source of [O$\,$III]
emission as long as $U$ is low enough in the midplane and the other
ratios can be matched.

We employ the shock models of Shull \& McKee (1979), as were also
considered by M97.  The line ratios in shock models, especially
[O$\,$III]/H$\beta$, are very sensitive to the shock velocity.  Other
variables include the preshock gas density and ionization state,
abundance, and transverse magnetic field strength.  The gas is assumed
to be initially neutral at $n_0=10$ cm$^{-3}$ and subsequently
penetrated by a precursor ionization front.  More appropriate to our
case would be a lower initial density (of order 0.1 cm$^{-3}$) and a
high initial ionization fraction.  We should also consider depleted
abundances to be consistent with the stellar ionization models, but SM
calculated only one such model to show the general effect of
depletions.  We do not perform an exhaustive search of parameter space
or carry out a statistical test of the goodness of fit, firstly
because an examination of figures such as Figure 9 can quickly reveal
which composite models are most successful, and secondly because some
of the fixed parameters are probably inappropriate for the halo of NGC
891 in any case.

In the composite models, we still consider that the stellar radiation
field is characterized by a decrease of $U$ with $z$, and that some
fraction of H$\alpha$ emission from shock ionization is added, with
this fraction possibly changing with $z$.  Although no model can
reproduce the line ratio behavior to within the errors, we find that
some of the main characteristics of the data can be reproduced.  One
of the most successful models is shown overlaid on the data in Figure
9 as the dashed lines joining open circles, which mark values of
log$\,U$.  The shock speed is 90 km s$^{-1}$.  At $z=0$ kpc, 7\%
of the H$\alpha$ emission arises from a 90 km s$^{-1}$ shock, rising
to 30\% by $z=2$ kpc.  The composite model is most successful if
log$\,U=-4.0$ at $z=0$ kpc, $-4.5$ at $z=1$ kpc (the limit of the
[O$\,$I]/H$\alpha$ data), and $-5.0$ at $z=2$ kpc (the limit of the
[O$\,$III]/H$\beta$ data).  Note that in the figure, circles indicate
log$\,U=$ --4.0, --4.3, --4.7 and --5.0, corresponding to $z=0-2$ kpc,
in all the panels despite the fact that the [S$\,$II]/[N$\,$II] ratio
includes data up to $z=3$ kpc and [O$\,$I]/H$\alpha$ is only detected
up to $z=1$ kpc.  Also shown in Figure 9 are the line ratios
for the shock models alone, namely, a 90 km s$^{-1}$ and a 100 km
s$^{-1}$ shock with standard abundances, and a 100 km s$^{-1}$ shock
with depleted abundances.

The major shortcoming of this composite model is in reproducing
[S$\,$II]/[N$\,$II] at low $z$.  For a given amount of [S$\,$II]
emission, the model overpredicts [N$\,$II].  At high $z$ this ratio is
somewhat more successfully modeled, but [N$\,$II]/H$\alpha$ simply
does not show as much disk-halo contrast as observed, as was the case
for the pure photo-ionization models.  Also, the predicted
[S$\,$II]/H$\alpha$ and [N$\,$II]/H$\alpha$ reach a plateau at 1.0 at
the lowest $U$ considered by S94, and thus cannot match the observed
continuing rise beyond $z=2$ kpc (Figure 9e).  In this regard
radiation-bounded models are more successful (S94), at least for
[S$\,$II]/H$\alpha$.  In that model, it is still rising at the
lowest considered $U$ value, so that its observed continued rise may
be explainable.  Finally, a low value of $U$ at $z=0$ pc is still
required, but again this may not be unreasonable given the arguments
mentioned above.

There is some freedom for variation of the shock parameters.  A 130 km
s$^{-1}$ shock (the highest speed considered by SM) contributing 2\%
of the H$\alpha$ emission at $z=0$ kpc, rising to 7\% at $z=2$ kpc,
produces almost identical curves in the diagnostic diagrams.  Since
only the [O$\,$III]/H$\beta$ ratio is large (7.35) for this model,
adding such a small amount of this emission affects mainly this ratio.
The composite model can be further explored by keeping the shock
contribution constant with $z$ but varying the shock speed.  A model
with a shock giving rise to 25\% of the H$\alpha$ emission, with a
speed of 60 km s$^{-1}$ at $z=0$ kpc, and 90 km s$^{-1}$ at $z=2$ kpc
(but with log$\,U=-4.7$ now) can reproduce the ranges of
[S$\,$II]/H$\alpha$, [N$\,$II]/H$\alpha$, [S$\,$II]/[N$\,$II], and
[O$\,$III]/H$\beta$ as successfully as the above models, but low-speed
shocks produce far too much [O$\,$I] emission.

SM find that the effect of introducing depleted abundances for a 100
km s$^{-1}$ shock is to raise [S$\,$II]/[N$\,$II], while
[O$\,$III]/H$\beta$ drops and [O$\,$I]/H$\alpha$ rises.  However,
assuming that the fractional changes also apply for a 90 km s$^{-1}$
shock, depleted abundances do not solve the [S$\,$II]/[N$\,$II]
problem, even if the composite model is started at a different value
of log$\,U$.  The effect on [S$\,$II]/[N$\,$II] is insufficient
because of the small contribution from shocks needed to fit
[O$\,$III]/H$\beta$.

Further latitude in the composite models may be gained by considering
a larger range of values for some of the other variables.  For
instance, Dopita \& Sutherland (1996) calculate properties of lower
density, magnetized shocks with speeds from 150 to 300 km s$^{-1}$.
They consider densities and magnetic field strengths obeying the
relation $0 < B/n_e^{1/2} (\mu $G cm$^{3/2}) < 4$.  These shocks have
strong radiative precursors which contribute a significant fraction of
the emission.  A model with shock speed 150 km s$^{-1}$, $n_e=1$
cm$^{-3}$ and no magnetic field produces line ratios similar to the 90
km s$^{-1}$ model considered above, with the exception that
[O$\,$I]/H$\alpha$ is about 80\% higher.  The main conclusion
from this subsection is that shocks are feasible as a secondary source
of energy input into the DIG of NGC 891, but it is difficult to
constrain their parameters with much confidence.

We have not speculated on the origin of the putative shocks.  But
given that the observed filaments suggest the presence of supershells
and chimneys, it is plausible that the shocks originate in such
expanding structures.  The shock speeds considered above are
reasonable when compared to superbubble calculations (e.g. MacLow,
McCray, \& Norman 1989).  In fact, the slit passes close to one of the
most prominent filaments, which may still have an associated shock.
This possibility highlights the importance of observing with many such
slit positions.

Finally, we point out that M97 also found that if a constant shock
speed model is used, then the fraction of the DIG emission that must
come from shocks increases with $z$.

\subsubsection{Composite Photo-ionization and Turbulent Mixing Layer Models}

Turbulent mixing layers (TMLs) are expected to occur at the interfaces
of hot and cold (or hot and warm) gas in the ISM of galaxies (Begelman
\& Fabian 1990; Slavin, Shull, \& Begelman 1993).  Superbubble walls
are a likely location for such layers.  Shear flows are expected along
the interface, leading to Kelvin-Helmholtz instabilities and
subsequent mixing of the gas.  The result is a layer of \lq\lq
intermediate temperature\rq\rq\ gas, probably at $5.0 \lesssim {\rm
log}\, T \lesssim 5.5$.  This gas may produce some fraction of the DIG
emission in galaxies.  Like typical DIG, it features enhanced
[S$\,$II]/H$\alpha$ and [N$\,$II]/H$\alpha$ relative to HII regions
and low [O$\,$I]/H$\alpha$.  However, unlike stellar-ionized gas,
[O$\,$III]/H$\beta$ should be of order 1--3.  Hence, TMLs provide a
source of enhanced [O$\,$III]/H$\beta$ and therefore may be relevant
to the current problem.

Slavin, Shull, \& Begelman (1993) calculate line ratios for TMLs as a
function of shear velocity, mixing layer temperature, initial cold
layer temperature, and abundance (solar vs. depleted).  We will
consider depleted abundances here.  Again, we will introduce a
contribution to the DIG emission from TMLs at some $z$ or $U$,
allowing for an increase in this contribution with $z$ as $U$
continues to decline, and consider only the PM model for the
photo-ionized component.  The best match to the data features a shear
velocity of 25 km s$^{-1}$ (the lowest modeled), a mixing layer
temperature of log$\, T = 5.3$, and an initially warm layer at 10$^4$
K rather than a cold layer.  This model is shown overlaid on the
diagnostic diagrams in Figure 10.  At $z=0$ kpc, 3\% of the H$\alpha$
emission arises from TMLs, rising to 15\% at $z=2$ kpc.  The rough
relation of log$\,U$ with $z$ is similar to the previous composite
model: log$\,U=-4.1$ at $z=0$ kpc, $-4.5$ at $z=1$ kpc (the limit of
the [O$\,$I]/H$\alpha$ data), and $-5.0$ at $z=2$ kpc (the limit of
the [O$\,$III]/H$\beta$ data).  Again, note that in the figure,
circles indicate log$\,U=$ --4.1, --4.3, --4.7 and --5.0,
corresponding to $z=0-2$ kpc, in all the panels despite the varying
maximum $z$ of the line ratio determinations.

The model is reminiscent of the best shock models and appears to be
somewhat more successful, but suffers from the same primary
shortcoming: the underprediction of [S$\,$II]/[N$\,$II] at low $z$.
Other TML models will not alleviate this problem because the model in
Figure 10 already features maximal [S$\,$II]/[N$\,$II].  As is the
case for shocks, there is room for flexibility in the parameters.  For
instance, [O$\,$III]/H$\beta$ is fairly constant in all the models
with depleted abundances, and all models regardless of abundance
feature very low [O$\,$I]/H$\alpha$.  [S$\,$II]/H$\alpha$ and
[N$\,$II]/H$\alpha$ show somewhat more variation with the input
parameters, but as most of the [S$\,$II] and [N$\,$II] emission arises
from the stellar-ionized gas, and the TML contribution is small, there
is reasonable latitude for variation of the parameters without
altering the resulting values of these ratios.  Again, the main point
is to demonstrate the feasibility of TMLs as a secondary source of
energy input rather than to find the best fitting parameters, or
indeed to show whether TMLs or shocks are preferred as the second
component.

\subsubsection{Other Possible Sources of Energy Input}

Shapiro \& Benjamin (1993) consider cooling, falling galactic fountain
gas initially raised from the midplane by supernovae.  The calculation
is followed from an initial temperature of 10$^6$ K to a final value
of 10$^4$ K.  While there are not yet predictions of optical emission
line ratios from such gas, one can expect [O$\,$III] emission as the
gas cools.  The run of [O$\,$III]/H$\beta$ with $z$ in a composite
model with fountain gas would depend on the fraction of diffuse
H$\alpha$ emission produced by the latter (the authors estimate that
it could account for perhaps 40\% in the Milky Way) and the details of
the cooling and dynamics, including the interaction with halo gas from
other processes.

Another idea that has not been deeply explored is heating of the halo
by microflares from magnetic reconnection events (Raymond 1992).  This
process may have a role in producing ultraviolet emission and
absorption lines, soft X-ray halo emission, and Reynolds layer
emission.  As the theory stands now, a broad range of
[O$\,$III]/H$\alpha$ values may result from this process, and there
are enough uncertainties as not to warrant a detailed comparison with
the data at this point.  The relevance of microflares should be
revealed with further refinements of the theory and additional
observations.

The problem of explaining low He$\,$I/H$\alpha$ in combination with
high [S$\,$II]/H$\alpha$ and [N$\,$II]/H$\alpha$ has motivated work on
other sources of heating for the DIG.  Since the forbidden lines are
highly temperature sensitive, small changes in temperature can be
important.  Minter \& Balser (1997) find that the dissipation of
turbulent energy in the ISM could raise the temperature of the DIG by
about 2000 K without additional ionization of helium, thus providing a
reasonable match to these line ratios in the Reynolds layer.  However,
there is still insignificant [O$\,$III] emission.  Photo-electric
heating from dust grains (Reynolds \& Cox 1992) should also have
little effect on [O$\,$III] emission.

Finally, scattered light from HII regions could be a source of
[O$\,$III] emission in the halo, but the rise in [O$\,$III]/H$\beta$
with $z$ would not be expected.

\section{Discussion and Conclusions}

Using a slit oriented parallel to, and offset 700 pc above, the major
axis of NGC 891, a spectrum of the DIG has been taken which reveals a
clear correlation of ratios of forbidden lines to H$\alpha$ with the
H$\alpha$ surface brightness.  The original motivation for this
observation was to search for variations in line ratios on and off the
filaments of DIG as further evidence that they are walls around
evacuated chimneys.  But although the filaments do show reduced
[S$\,$II]/H$\alpha$, [N$\,$II]/H$\alpha$, and [O$\,$I]/H$\alpha$
relative to gas on adjacent lines of sight, the contrast merely
reflects the overall correlation.  The relationship probably indicates
that regions of brighter H$\alpha$ emission receive a radiation field
with a higher ionization parameter.  Also, although some of the
filaments show deviations from the observed smooth trend of mean
velocity with position along the major axis, these departures could
simply be due to geometric effects: if the filaments are inner galaxy
features, they will bias the mean velocity for their line of sight
away from the systemic velocity.  [S$\,$II]/[N$\,$II] surprisingly
shows no significant variation along the slit.  Finally, the
H$\alpha$-emitting halo gas at this height is about 80--95\% ionized,
based on the observed range of [O$\,$I]/H$\alpha$ and assuming
$T=10^4$ K.  The correlation of [O$\,$I]/H$\alpha$ with surface
brightness probably reflects a higher degree of ionization where the
photon field is more intense.

Results from this observation emphasize the difficulty in interpreting
DIG observations of edge-on galaxies.  Confusion is caused by
uncertainties in the location of a parcel of gas along the line of
sight, its effective distance from a source of ionization and other
unrelated gas in the same direction.  It is difficult from such
observations to draw conclusions about the environment of the
filament, for example.

Spectra from a slit oriented perpendicular to the plane at $R=100''$
along the major axis on the NE side show a rise of
[S$\,$II]/H$\alpha$, [N$\,$II]/H$\alpha$, and [O$\,$I]/H$\alpha$ with
$z$.  At the midplane, [O$\,$I]/H$\alpha$ values indicate that H is
essentially 100\% ionized, dropping to 90\% at $z=1$ kpc, assuming
$T=10^4\,$K.  The $z$-dependence of these line ratios is expected if
the gas is ionized by massive stars in the disk.  However, it is
unexpectedly found that [O$\,$III]/H$\beta$ also rises with $z$,
whereas it should decline with $z$ in photo-ionization models.  This
result necessitates the consideration of secondary sources of
ionization.  [S$\,$II]/[N$\,$II] unexpectedly shows essentially no
dependence on $z$ and H$\alpha$ surface brightness.  Put another way,
[N$\,$II]/H$\alpha$ shows the same disk-halo contrast as that of
[S$\,$II]/H$\alpha$, whereas a smaller contrast is expected.

Strong [O$\,$III] emission is expected from several energetic procesess.
We considered shocks and turbulent mixing
layers as sources of such gas.  Models in which a small fraction of
the H$\alpha$ emission comes from one of these mechanisms can be made
to fit the data reasonably well, but most noticeably the remarkable
constancy of [S$\,$II]/[N$\,$II] with $z$ is still difficult to
reproduce.  In the case of shocks, it is difficult to constrain the
shock speed or the contributed fraction of the DIG emission at this
point.  Of course, the line of sight may sample shocks with a range of
speeds and some mean value.  There is also significant latitude in the
parameters in the case of TMLs.  Other sources of strong [O$\,$III]
emission may include cooling galactic fountain gas and microflares
from magnetic reconnection.  These should be explored further in light
of the current results.  Because of these facts and possibilities, the
results are meant only to indicate the feasibility of such classes of
models and the likelihood that one or more physical processes is
producing intermediate temperature gas in the halo of NGC 891.

On the other hand, the finding that the second source of line emission
becomes more important as $z$ increases may be reasonable.  In the
case of TMLs, for example, Shull \& Slavin (1994) point out that this
process may indeed be more common at large $z$, where superbubbles
break out of the thin disk gas layers, producing Rayleigh-Taylor
instabilities and shear flows that lead to the mixing.  These authors
were attempting to explain the larger scale-height of C$\,$IV UV
absorption line gas relative to N$\,$V (Sembach \& Savage 1992) as an
increasing predominance of TMLs over SN bubbles with height off the
plane -- the former producing higher C$\,$IV/N$\,$V.  If the TML
process begins only at the approximate height where breakout occurs,
while the stellar radiation field is increasingly diluted with $z$,
then an increasing fraction of H$\alpha$ emission from TMLs may be
quite reasonable.  The rough fractions found in \S3.2.3 are comparable
to those expected by Slavin et al. (1993) for the Milky Way diffuse
H$\alpha$ emission.

If the photo-ionized and secondary components of the DIG emission both
arise in exponential layers with different scale-heights, then the
composite models, although illustrative, can be used to estimate
roughly the relative scale-heights.  In the two shock models and one
TML model considered, the fraction of emission arising from the second
component is 3--5 times higher at $z=2$ kpc than at $z=0$ kpc.
Assuming exponential layers, the scale-height of the second component
must be 3--4 times that of the photo-ionized component.  This
conclusion is very tentative, however, given the uncertainties in the
modeling and the lack of information on [O$\,$III]/H$\beta$ at higher
$z$.  It it is tempting to identify the second component with the
high-$z$ tail of H$\alpha$ emission found in the deeper spectrum of Rand
(1997a).  However, the exponential scale-height of this tail is 5--7
times that of the main component, and it contributes about 50\% of the
emission at $z=2$ kpc.  While these numbers do not quite match those
for the second component proposed here, there may yet prove to
be a connection.

For the photo-ionized component, we found the most success by using
the model from S94 with the lowest terminal hydrogen column considered
for individual clouds in the DIG layer.  Using larger columns tends to
push the model curves towards those of the radiation-bounded case,
which is found to be very difficult to incorporate in a successful
composite model.  This constraint suggests that the DIG consists of
quite small clumps (or filaments or sheets, since S94's calculation is
one-dimensional) of several pc thickness, for a representative density
of $n_e=0.1$ cm$^{-3}$.  If this conclusion is not borne out by future
observations, the composite models presented here will need to be
reconsidered.  The S94 model used here also features the hardest
emergent stellar spectrum considered (in order to produce high
[S$\,$II]/H$\alpha$ and [N$\,$II]/H$\alpha$), but more modest spectra
may be allowable if other sources of non-ionization heating are at
work (e.g. Minter \& Balser 1997).

Despite complications introduced by the second DIG component, it
should be noted that the observed properties of the three ratios
[S$\,$II]/H$\alpha$, [N$\,$II]/H$\alpha$, and [O$\,$I]/H$\alpha$ with
$z$ are still reasonably explained {\it to first order} by
photo-ionization models alone.  Their smooth increase with $z$ is as
predicted as are their rough values.  Emission from the second
component probably has only a secondary effect on these ratios.  The
$z$-independence of [S$\,$II]/[N$\,$II] is not understood in either a
pure photo-ionization or composite model.

An undesirable aspect of the composite models considered here is that
emission from photo-ionized and (for example) shock ionized gas is
simply added together with no unified physical picture in mind.  It
would be desirable eventually to have, say, a calculation of the
evolution of a superbubble which included photo-ionization, shocks and
TMLs in a more self-consistent way.  For instance, what is the effect
of the radiative precursor of a shock which enters gas already ionized
by dilute stellar radiation?

The emission line properties revealed by the perpendicular slit share
some similarities with those of the halo of the starburst galaxy M82.
In the Fabry-Perot data of Shopbell \& Bland-Hawthorn (1997),
[N$\,$II]/H$\alpha$ shows a general tendency to rise with $z$ on the
N side, up to the limit of measurability at about $z$=750 pc.  On
the S side, there is little dependence on z, with perhaps 0.6
typical.  [O$\,$III]/H$\alpha$=0.03 at $z=0$ pc and 0.08 at $z=750$ pc
(values of [O$\,$III]/H$\beta$ are about 3 times higher assuming
little extinction).  In a long-slit spectrum through the halo of M82,
M97 sees higher values of [O$\,$III]/H$\beta$, reaching 0.7 at $z=1$
kpc.  The sequence of points in her diagnostic diagram of
[S$\,$II]/H$\alpha$ vs. [O$\,$III]/H$\beta$ has a rising slope,
as in NGC 891, although much steeper (other irregulars show a falling
or nearly flat slope).  [O$\,$I]/H$\alpha$ also rises with $z$ in M82,
and shows a range of values (about 0.01 to 0.1) similar to those
reported here.  Shopbell \& Bland-Hawthorn (1997) point out that their
ratios become more shock-like with distance from the starburst, as also
noted by Heckman, Armus, \& Miley (1990).  This behavior is now seen
in a spiral halo as well.

Other evidence for multi-phase halos is provided by the study of NGC
4631 by Martin \& Kern (1998).  They detect an extensive halo of
[O$\,$III] emission which spatially coexists with the observed soft
X-ray and H$\alpha$ emitting halos.  Within this halo are several
bright [O$\,$III] condensations in which the measured
[O$\,$III]/H$\beta$ ratio is $\sim 20$.  On this basis, they argue
that the H$\alpha$ and [O$\,$III] emission is tracing distinct
components in a multi-phase halo medium.

It should be pointed out that [O$\,$III]/H$\alpha$ values of
0.12--0.21 (comparable to values in NGC 891) are seen in the DIG of
M31 (Greenawalt et al. 1997).  There is little
correlation of this ratio with the brightness of the DIG or its
remoteness from visible HII regions.  The high [O$\,$III]/H$\alpha$ in
this case may be due to a radiation field not as dilute as in the halo
of NGC 891 (for instance, [N$\,$II]/H$\alpha$ and [S$\,$II]/H$\alpha$
are substantially lower than in the halo of NGC 891 at $z=1-2$ kpc).
Greenawalt et al. (1997) also find that TMLs may contribute some
fraction of the H$\alpha$ emission in regions of very faint DIG, but
no more than about 20\% and most likely only a few percent, similar
to the findings for NGC 891.

More light has been shed on the question of [O$\,$III]/H$\beta$ trends
in the DIG of face-ons by Wang et al. (1997).  For three
of their five galaxies with good [O$\,$III] detections in their DIG
spectra, they find that [O$\,$III]/H$\beta$ is in the range $<0.2$ to
2 and, for a given slit, is systematically higher than in the HII
regions in that slit.  They also find that the [O$\,$III] line widths
are usually larger than those of [N$\,$II], and thus refer to a \lq\lq
quiescent\rq\rq\ photo-ionized DIG component which accounts for the
bulk of the H$\alpha$, [N$\,$II] and [S$\,$II] emission, and a \lq\lq
disturbed\rq\rq\ component (shocks and TMLs are considered)
contributing a minority ($<$ 20\%) of the H$\alpha$ emission but
responsible for the [O$\,$III] emission.  For vertical hydrostatic
equilibrium, the contrast in line-widths indicates that the
scale-height of the disturbed component is 1.5--2 times greater than
that of the quiescent DIG.  In NGC 891, the line-widths are dominated
by galactic rotation and thus such an analysis cannot be carried out.

The existence of a second source of line emission may be relevant for
the issue of the low He$\,$I/H$\alpha$ ratio.  The value of 0.027 for
the lower halo of NGC 891 implies a much softer spectrum than is
required to explain the high [N$\,$II]/H$\alpha$ and
[S$\,$II]/H$\alpha$ (R97).  Apart from the possibility that the
forbidden line emission is complicated by additional sources of
ionization and non-ionizing heating, He$\,$I/H$\alpha$ may also be
affected by secondary ionization sources if they contribute sufficient
H$\alpha$ emission.  For instance, the ratio is fairly sensitive to
shock conditions.  The SM 100 km s$^{-1}$ model
gives a value of 0.027.  The ratio is 0.005 for an 80 km s$^{-1}$
shock, but this model predicts insignificant [O$\,$III] emission.  The
Dopita \& Sutherland (1996) higher velocity, lower density (n=1),
magnetized models give a much more significant ionized precursor.  For
the lowest velocity considered, 150 km s$^{-1}$, the shock itself
gives a low ratio of 0.019, but the line fluxes for the precursor are
not given.  For a 200 km s$^{-1}$ shock, the ratio from combined shock
and precursor is 0.046.  Slavin, Shull, \& Begelman (1993) do not
predict He$\,$I emission.  Regardless, if shocks or TMLs can provide,
say 25\% of the H$\alpha$ emission at $z=2$ kpc, then there may be a
region of parameter space which can produce low enough
He$\,$I/H$\alpha$ so that the composite line ratio is significantly
reduced below that of the stellar-ionized gas alone.  The
contributions of these sources required in \S 3 may not be
sufficiently large, but the effect is worth future consideration.  The
inferred stellar temperature, mean spectral type and upper IMF cutoff
would then all be underestimated.

Finally, it is worth re-emphasizing that all emission line fluxes and
ratios presented here are averaged along a line of sight through the
DIG layer, and that local variations in, for example, the derived
ionization fraction of H surely exist.  Also, although the vertical
dependence of the line ratios has been very revealing, only one slit
position has been observed, covering the halo above the most active
region of star formation in the disk, and close to an H$\alpha$
filament.  A key question is how these halo properties vary with
environment.  Does [O$\,$III]/H$\beta$ show the same behavior above
more quiescent parts of the disk?  Is this behavior peculiar to the
halo of NGC 891 only, or is it a general feature of DIG halos?  These
questions will be addressed by further observations.

\vspace*{0.5in}

The author has benefited from many useful discussions about DIG
ionization from R. Reynolds, R. Walterbos (whose comments as referee
also improved the paper), J. Slavin, R. Benjamin, J. Shields, and
others.  The help of the KPNO staff is also greatly appreciated.

\newpage

\figcaption{The slit positions for the two observations are shown
overlaid on the H$\alpha$ image of NGC 891 from Rand, Kulkarni, \&
Hester (1990) in (a), and on a version of the central part of the same
image in (b), processed to show how the \lq\lq parallel\rq\rq\ slit
traverses filaments of H$\alpha$ emission (see text).\label{fig1}}

\figcaption{Dependence of line ratios on slit position for the
parallel slit.  In each panel, the normalized H$\alpha$ surface
brightness is plotted.  Positions of the four brightest filaments are
indicated with arrows in (a).  The line ratios plotted are (a)
[S$\,$II] $\lambda 6716$/H$\alpha$, (b) [N$\,$II]
$\lambda$6583/H$\alpha$, (c) [O$\,$I] $\lambda 6300$/H$\alpha$, and
(d) [S$\,$II] $\lambda$6716/[N$\,$II] $\lambda$6583.\label{fig2}}

\figcaption{The dependence of (a) [N$\,$II] $\lambda$6583/H$\alpha$
(filled symbols), [S$\,$II] $\lambda 6716$/H$\alpha$ (open symbols),
and (b) 3$\times$[O$\,$I] $\lambda 6300$/H$\alpha$ (filled symbols)
and [S$\,$II] $\lambda 6716$/H$\alpha$/[N$\,$II]
$\lambda$6583/H$\alpha$ (open symbols) on H$\alpha$ surface
brightness.  Triangles indicate data points at locations of filaments
marked in Figure 2.  Note that the [O$\,$I]/H$\alpha$ values and error
bars have been multiplied by 3.\label{fig3}}

\figcaption{The intensity-weighted mean heliocentric velocity
vs. location determined from the H$\alpha$, [S$\,$II], and [N$\,$II]
lines.  Note some deviations from the general trend at the positions
of filaments, which are indicated by arrows.  Velocity uncertainties
generally increase with decreasing intensity and are $1-2$ km s$^{-1}$
for all points except those to the SW of the lowest intensity
filament, where they rise to $3-5$ km s$^{-1}$.\label{fig4}}

\figcaption{Spectrum in the region of line splitting showing the
H$\alpha$ and two [N$\,$II] lines.  The spectrum is an average
over 5 pixels (3.4'' or 150 pc) at $R=4$ kpc on the SE side.\label{fig5}}

\figcaption{Dependence of line ratios on slit position for the
perpendicular slit.  In each panel, the normalized H$\alpha$ surface
brightness is plotted.  At $z \lesssim 1$ kpc, dust absorption causes
an apparent local minimum in the H$\alpha$ emission. The line ratios
plotted are (a) [S$\,$II] $\lambda 6716$/H$\alpha$, (b)
[N$\,$II] $\lambda $6583/H$\alpha$, (c) [O$\,$I] $\lambda
6300$/H$\alpha$, (d) [S$\,$II] $\lambda 6716$/[N$\,$II] $\lambda$6583,
and (e) [O$\,$III] $\lambda 5007$/H$\beta$.\label{fig6}}

\figcaption{The dependence of (a) [N$\,$II] $\lambda 6583$/H$\alpha$
(filled circles), [S$\,$II] $\lambda 6716$/H$\alpha$ (open circles),
[S$\,$II] $\lambda 6716$/[N$\,$II] $\lambda 6583$ (crosses),
and (b) 3$\times$[O$\,$I] $\lambda 6300$/H$\alpha$ (filled circles)
and [O$\,$III] $\lambda 5007$/H$\beta$ (open circles) on H$\alpha$
surface brightness.  Note that the [O$\,$I]/H$\alpha$ line ratio
values and error bars have been multiplied by 3.\label{fig7}}

\figcaption{Diagnostic diagrams of logarithmic line ratios: (a)
[O$\,$III]/H$\beta$ vs.  [N$\,$II]/H$\alpha$, (b) [O$\,$III]/H$\beta$
vs.  [S$\,$II]/H$\alpha$, (c) [O$\,$I]/H$\alpha$
vs. [N$\,$II]/H$\alpha$, (d) [O$\,$I]/H$\alpha$
vs. [S$\,$II]/H$\alpha$, and (e) [S$\,$II]/H$\alpha$
vs. [N$\,$II]/H$\alpha$.  The curves are predictions from the
radiation-bounded dilute photo-ionization model (called PM in the
text) from S94.  Open circles indicate values of the logarithm of the
ionization parameter, log$\,U=-2.3,\ -2.7,\ -3.0,\ -3.3,\ -3.7,\
-4.0,\ -4.3,$ and $-4.7$.  In (e), the sequence begins at
log$\,U=-2.7$.\label{fig8}}

\figcaption{Diagnostic diagrams of logarithmic line ratios: (a)
[O$\,$III]/H$\beta$ vs.  [N$\,$II]/H$\alpha$, (b) [O$\,$III]/H$\beta$
vs.  [S$\,$II]/H$\alpha$, (c) [O$\,$I]/H$\alpha$
vs. [N$\,$II]/H$\alpha$, (d) [O$\,$I]/H$\alpha$
vs. [S$\,$II]/H$\alpha$, and (e) [S$\,$II]/H$\alpha$
vs. [N$\,$II]/H$\alpha$.  The curves are predictions from a
matter-bounded dilute photo-ionization model (called PM in the text)
from S94.  Small open circles indicate values of the logarithm of the
ionization parameter, log$\,U=-3.3,\ -3.7,\ -4.0,\ -4.3,\ -4.7$ and
$-5.0$, although the first value may differ between plots.  The
triangle, square, and pentagon represent, respectively, line ratios
from the standard-abundance 90 km s$^{-1}$ and 100 km s$^{-1}$ shock
models and the depleted-abundance 100 km s$^{-1}$ shock model of SM.
The large open circles joined by dashed lines represent the composite
photo-ionization/shock-ionization model.  These circles represent
values of the ionization parameter of the photo-ionized component,
log$\,U=-4.0,\ -4.3,\ -4.7$ and $-5.0$.  The fraction of emission from
the 90 km s$^{-1}$ shock ranges from 7\% at the left end of the sequence
to 30\% at the right end.  Note that the lowest appropriate value of
log$\,U$ for a given line ratio depends on the maximum height above
the plane at which it has been detected.  See text for the best values
of log$\,U$ at various heights.\label{fig9}}

\figcaption{Diagnostic diagrams as in Figure 9 except that the
composite photo-ionization/turbulent mixing layer model is now plotted
as the large open circles joined by dashed lines.  These circles
represent values of the ionization parameter of the photo-ionized
component, log$\,U=-4.1,\ -4.3,\ -4.7$ and $-5.0$.  The fraction of
emission from the TML model ranges from 3\% at the left end of the
sequence to 15\% at the right end.  Note that the lowest appropriate value
of log$\,U$ for a given line ratio depends on the maximum height above
the plane at which it has been detected.  See text for the best values
of log$\,U$ at various heights.  The triangle represents line ratios
from a pure TML model.\label{fig10}}

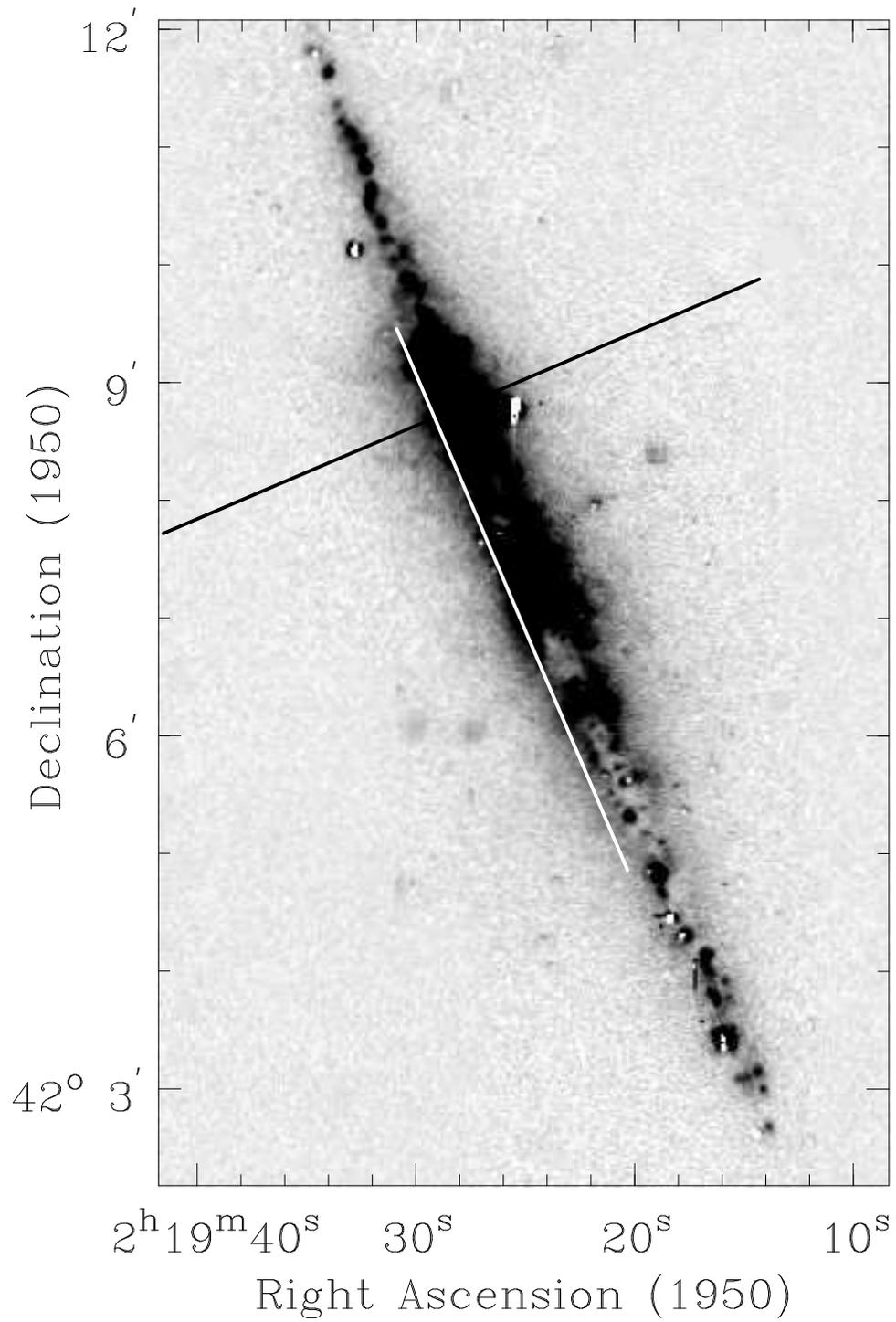
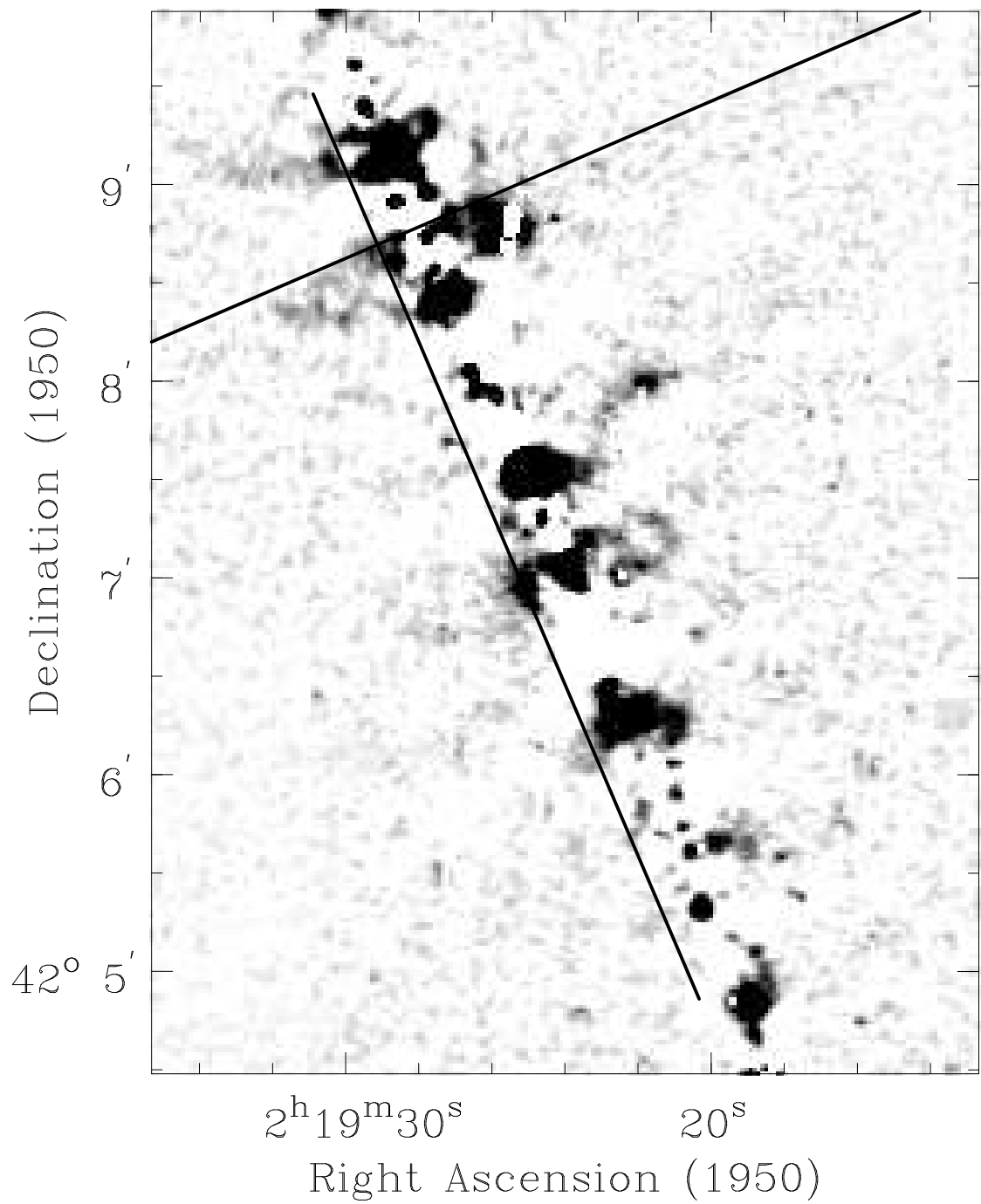
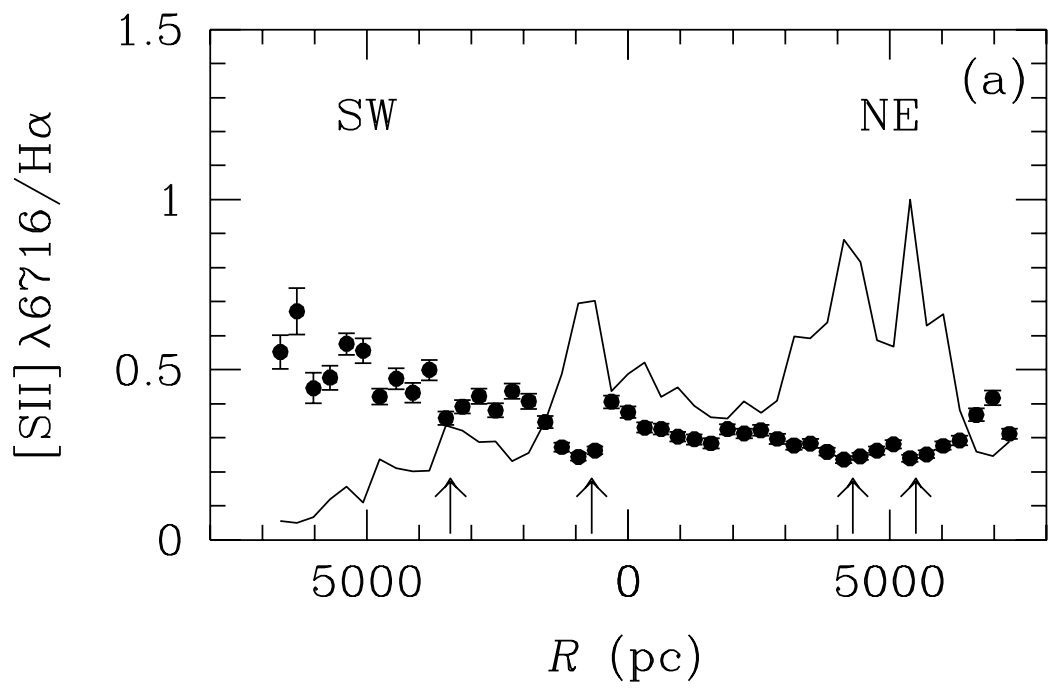
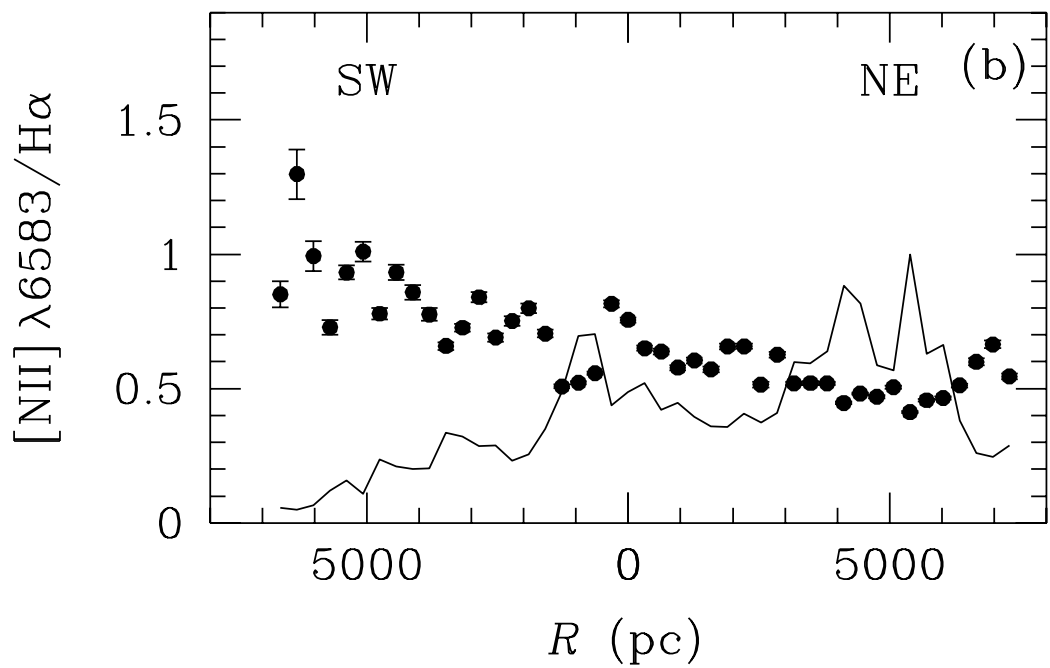
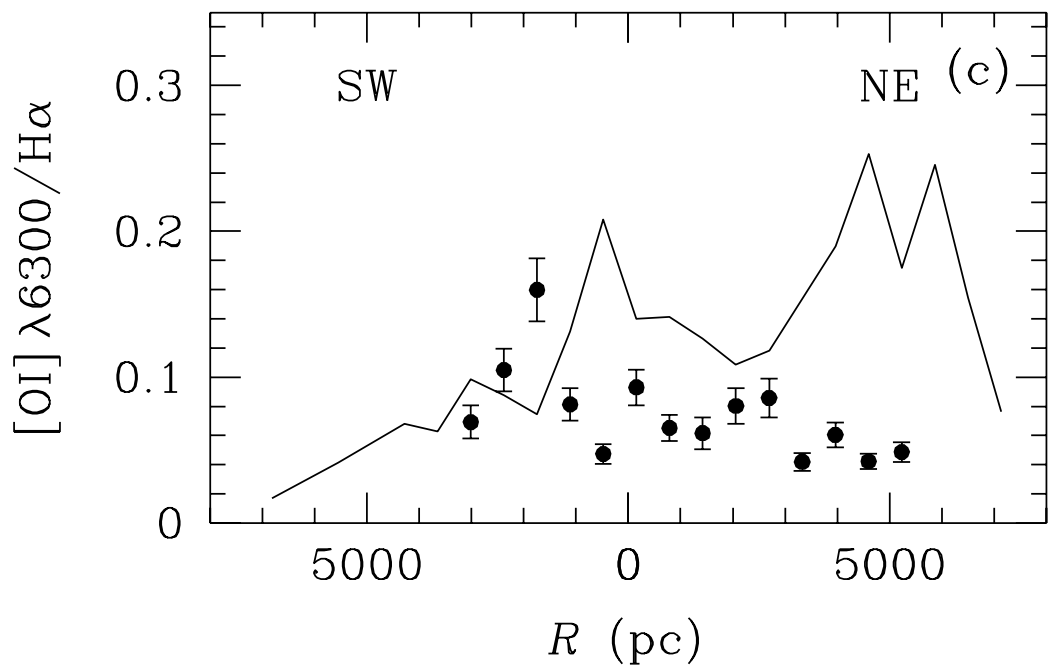
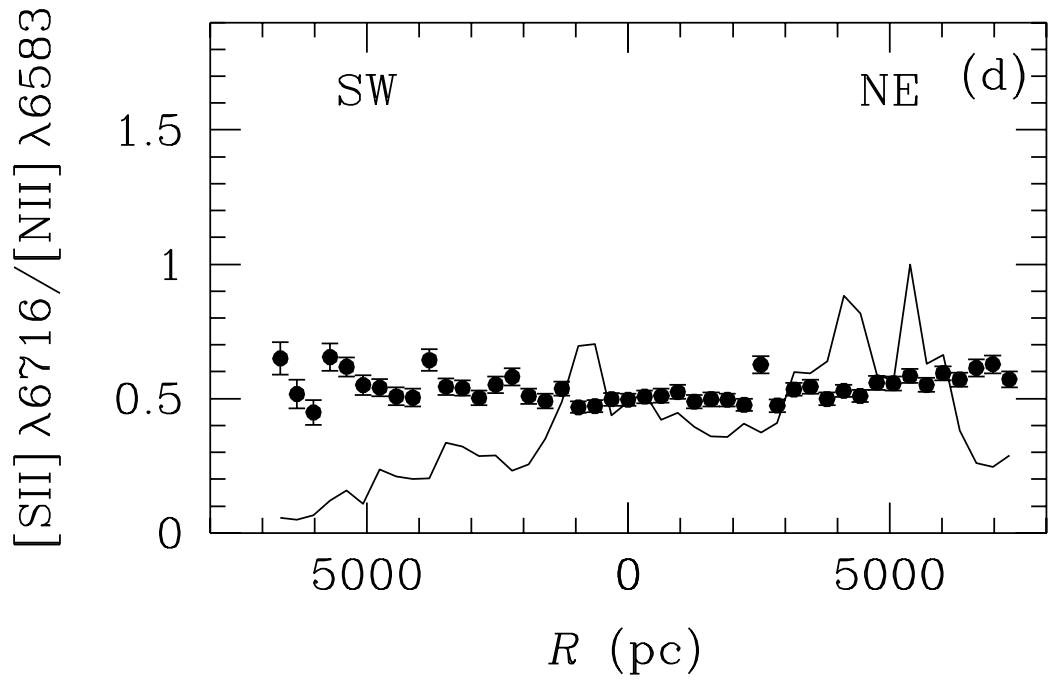
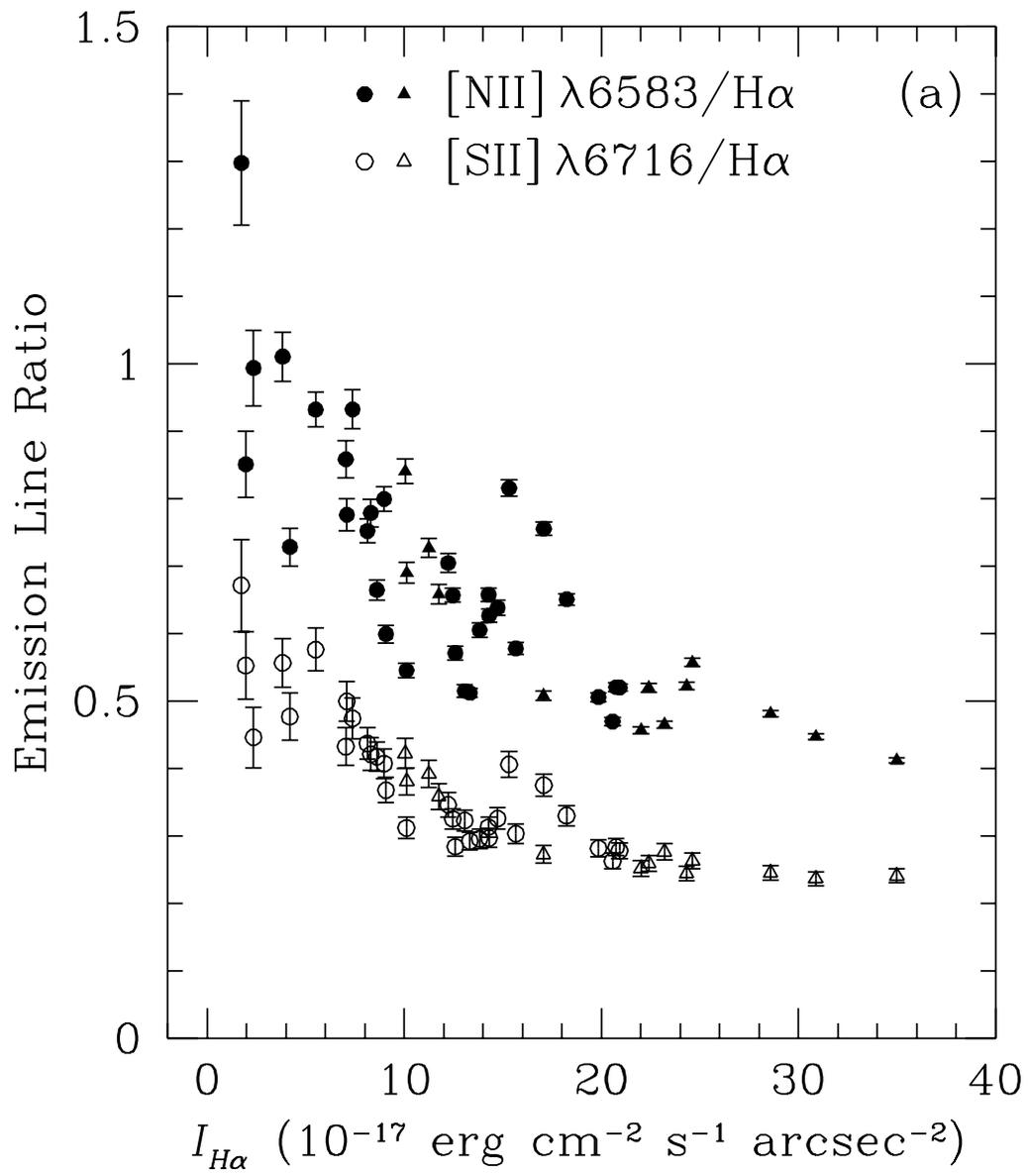
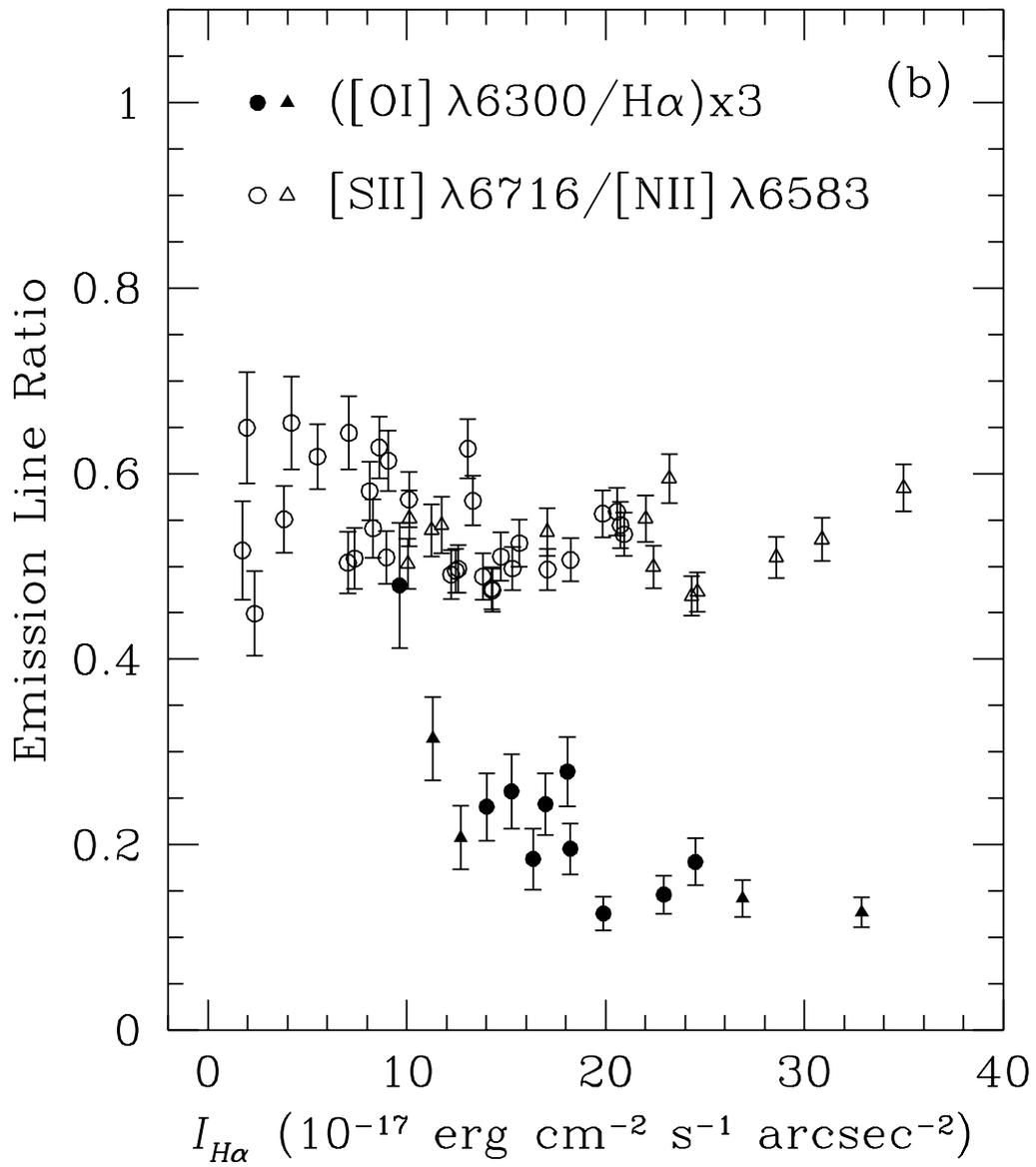
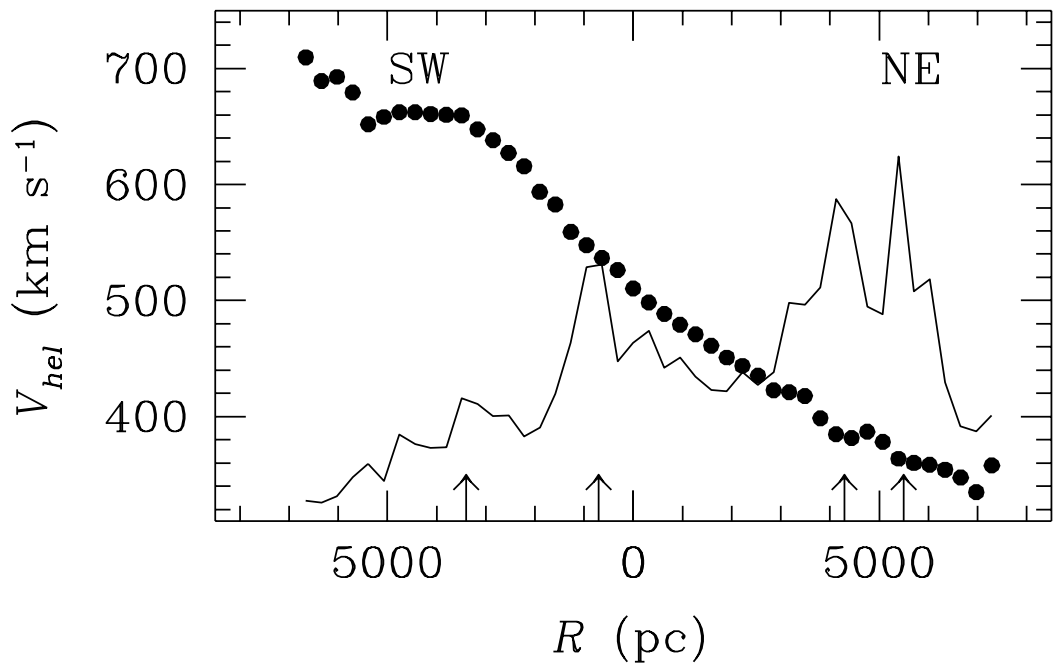
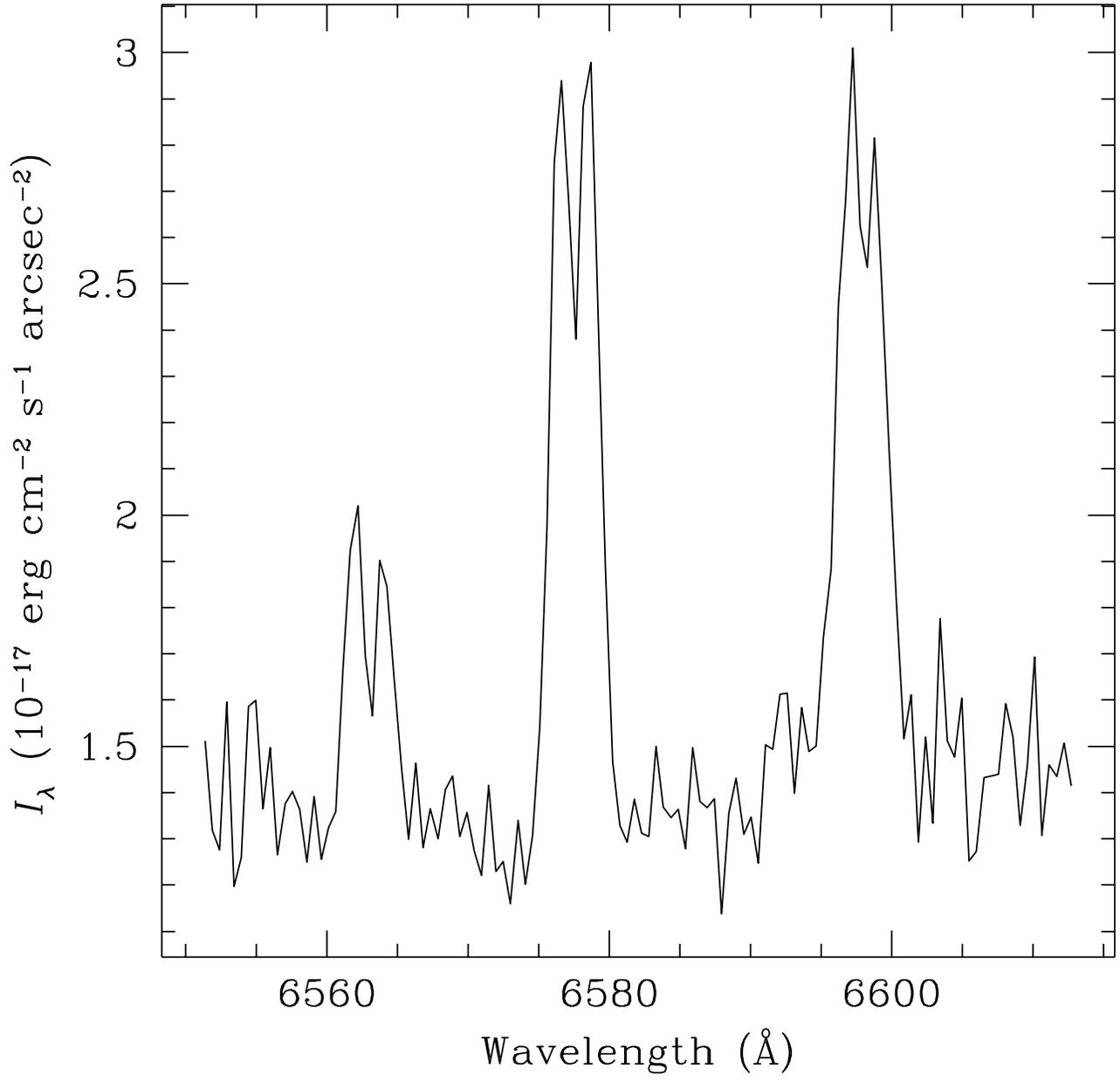
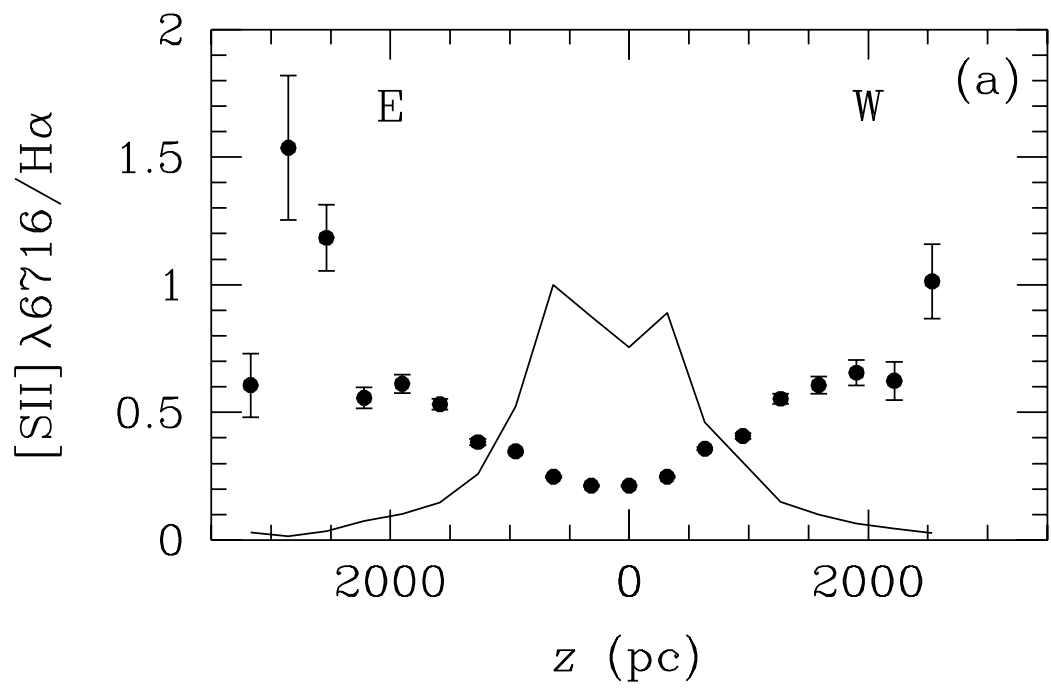
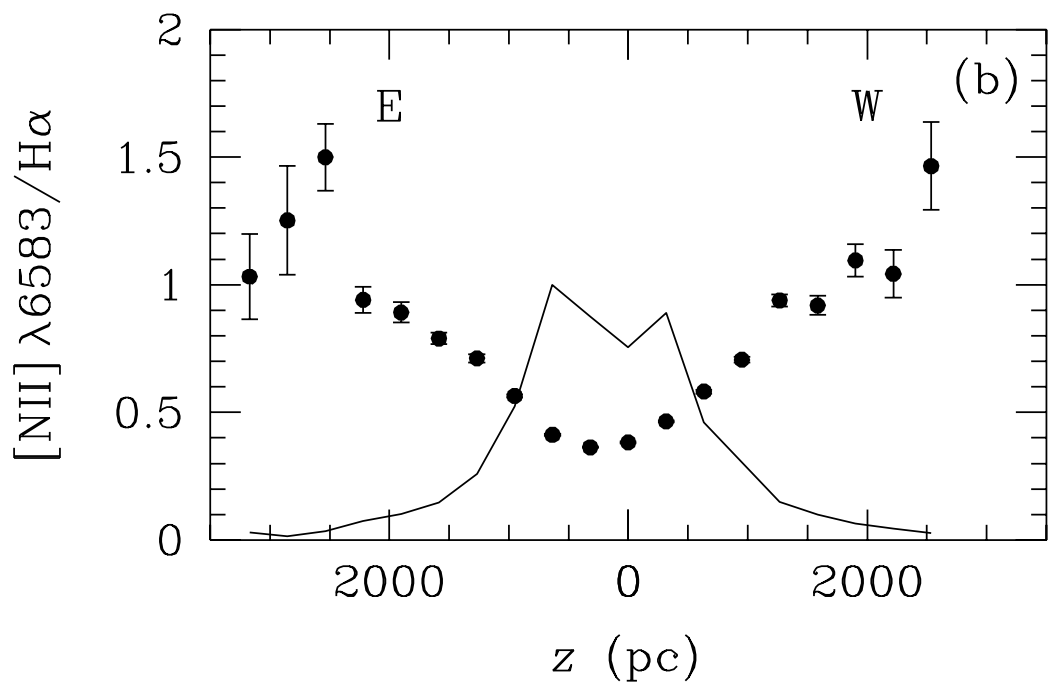
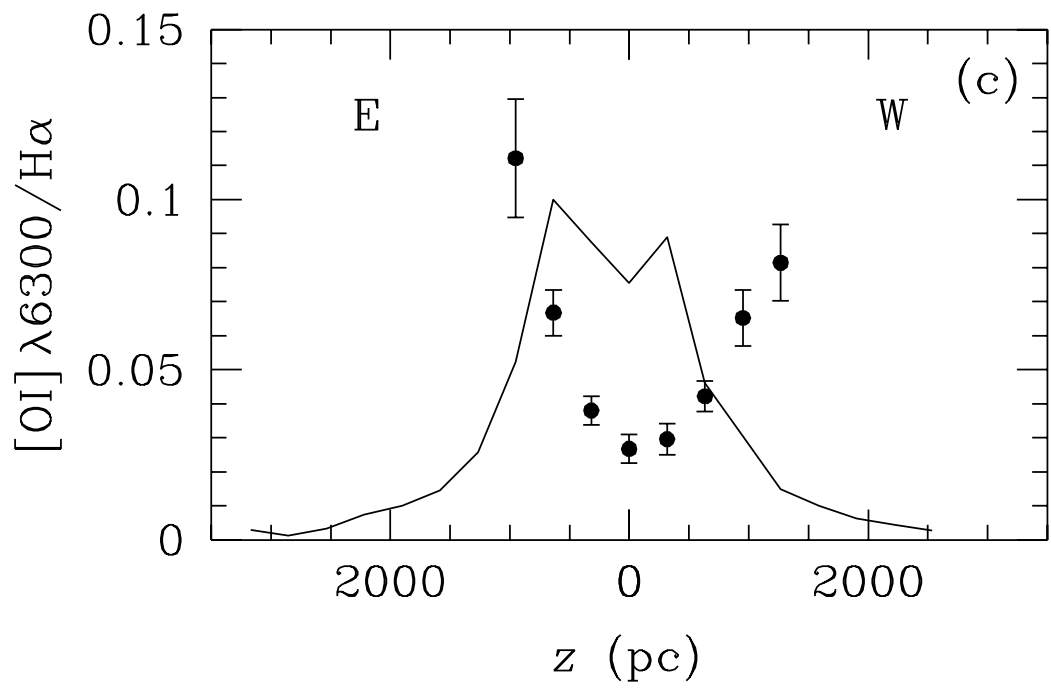
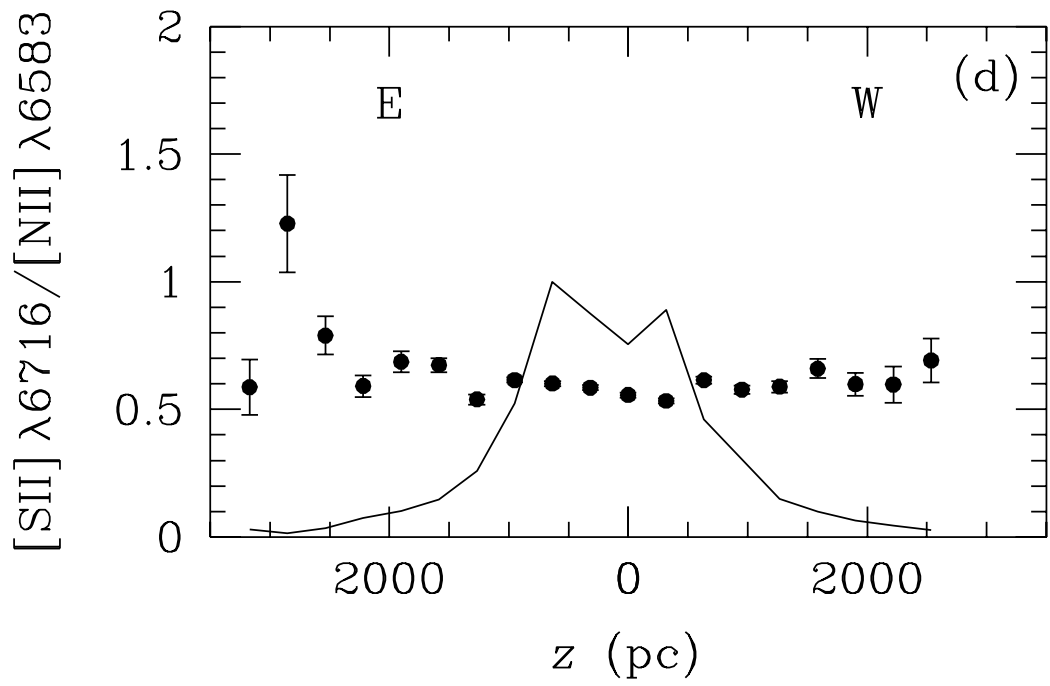
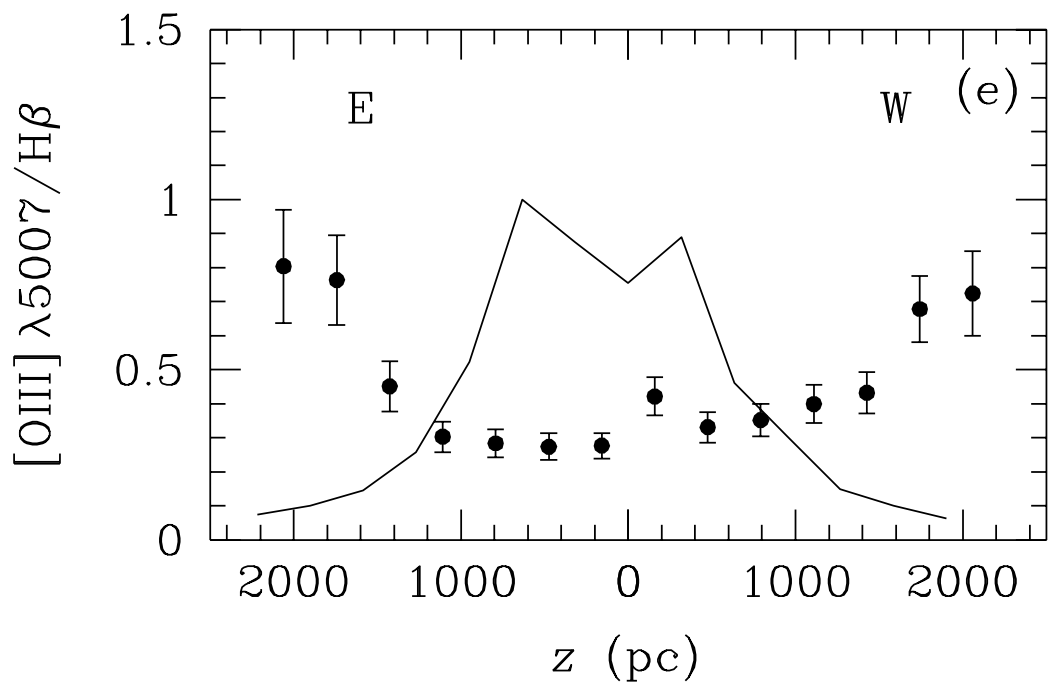
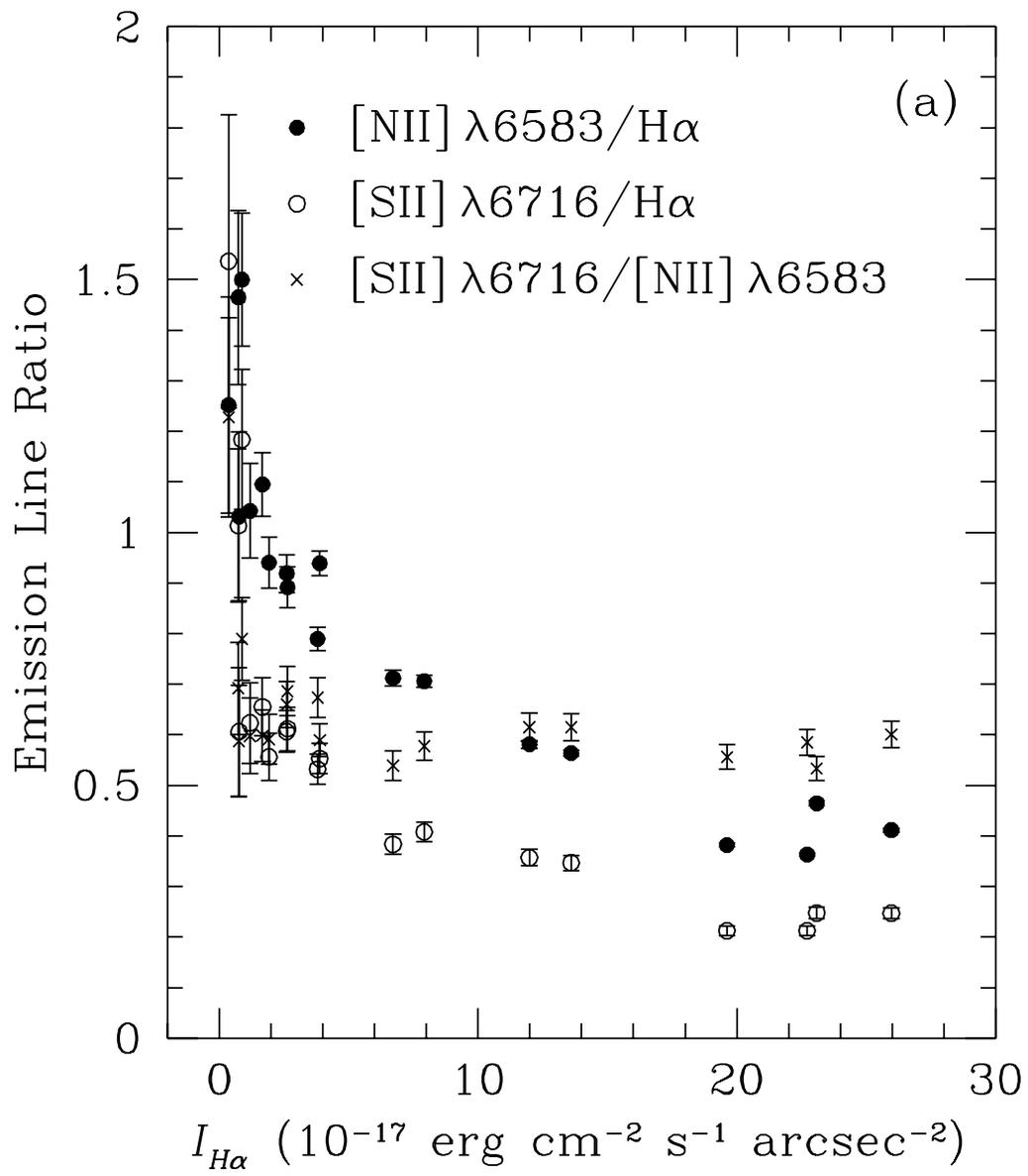
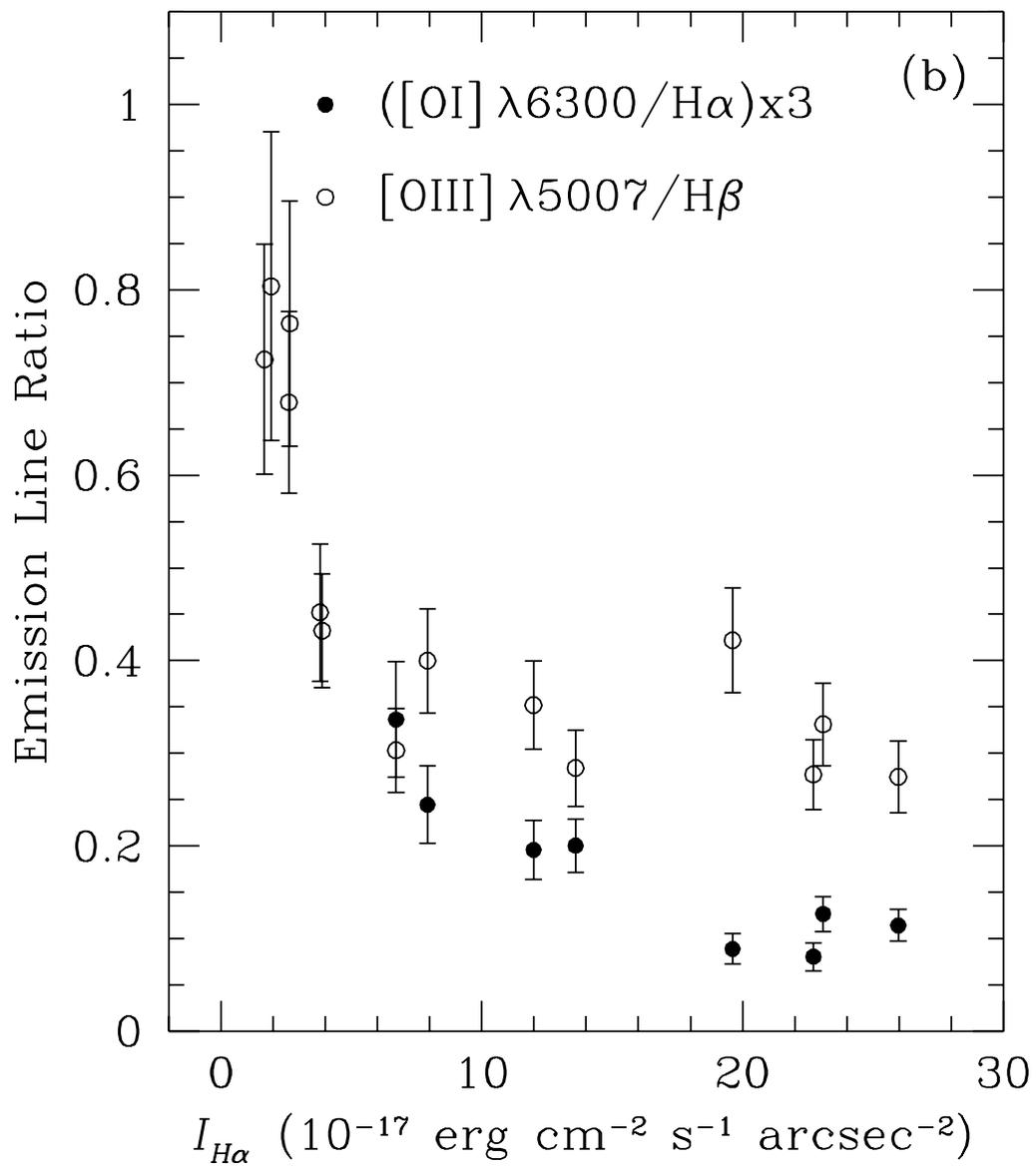
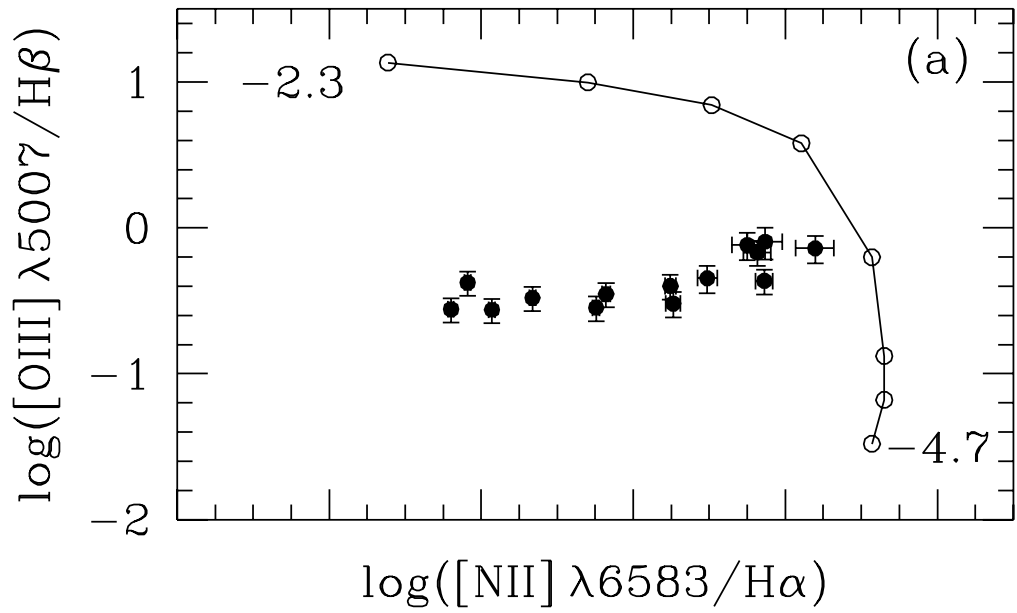
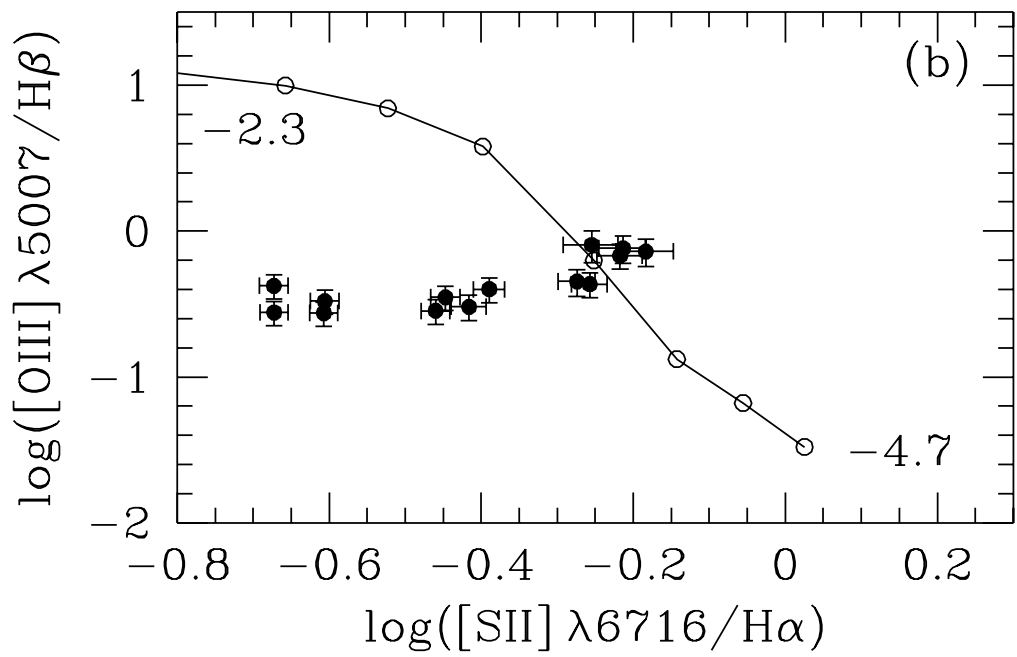
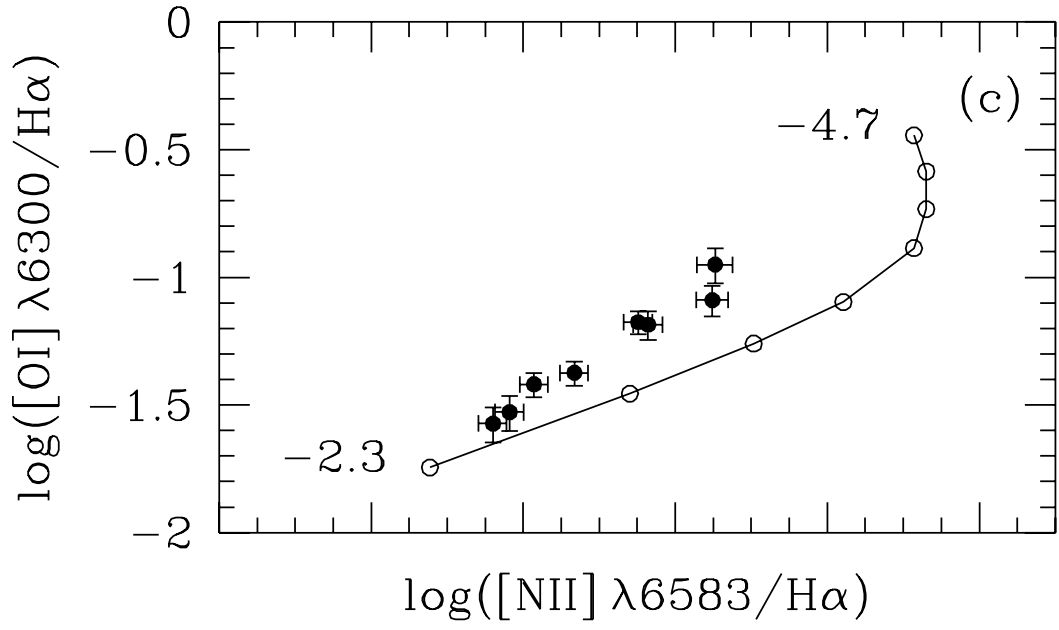
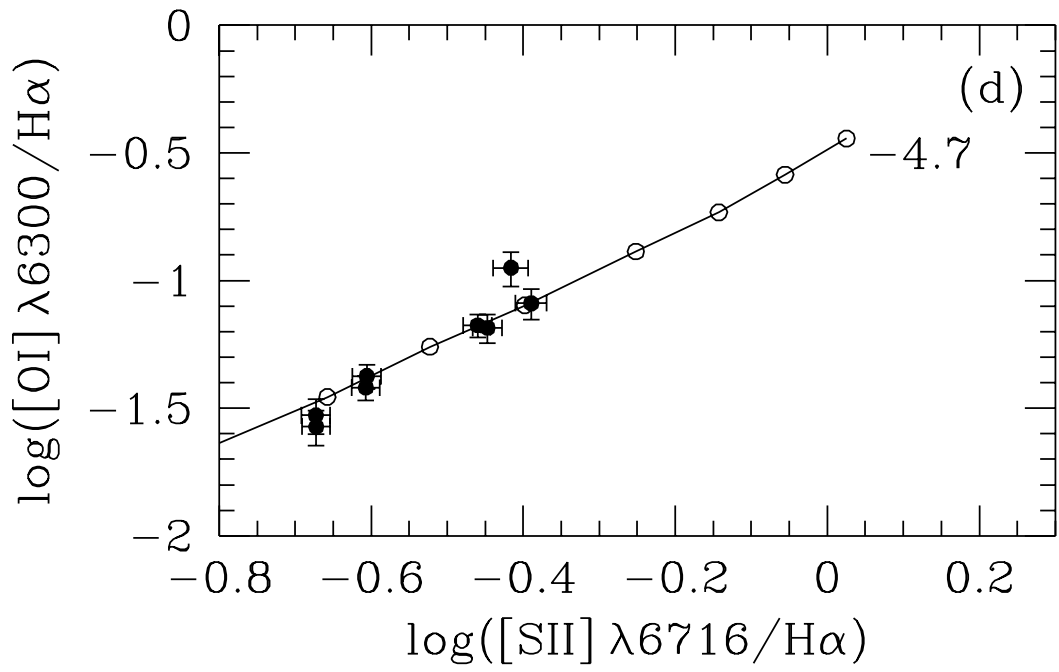
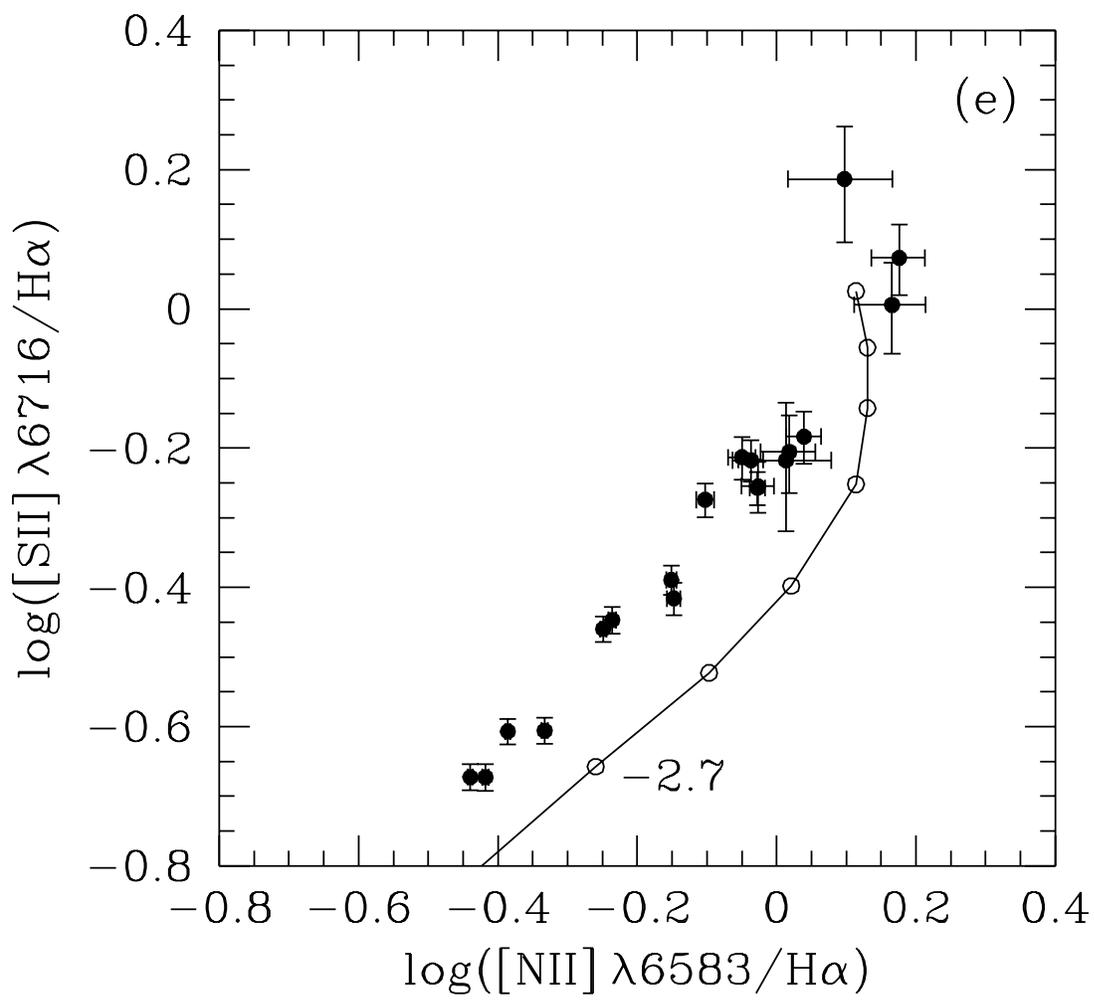
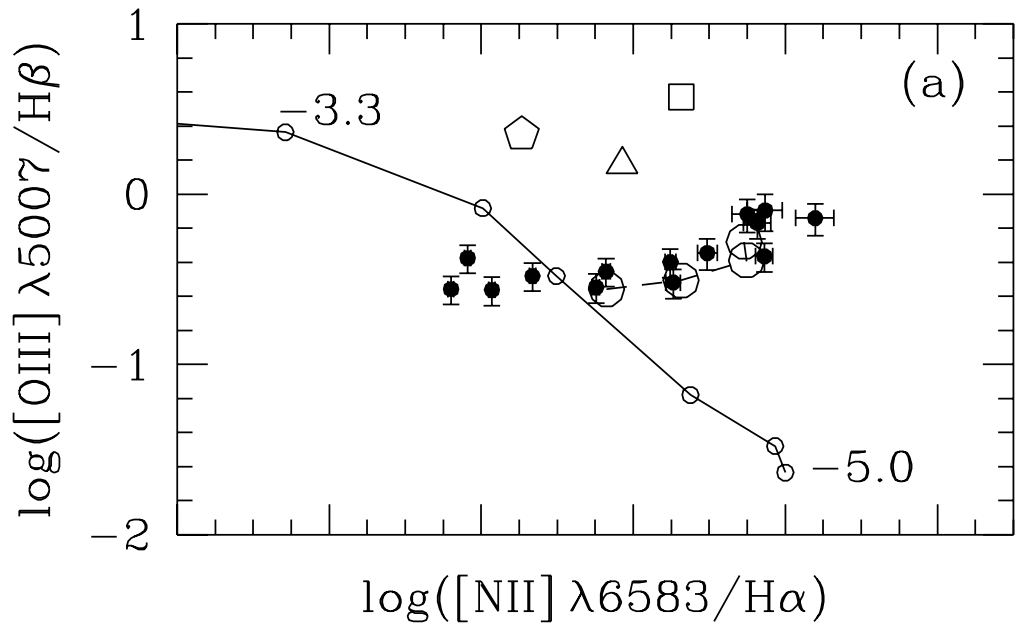
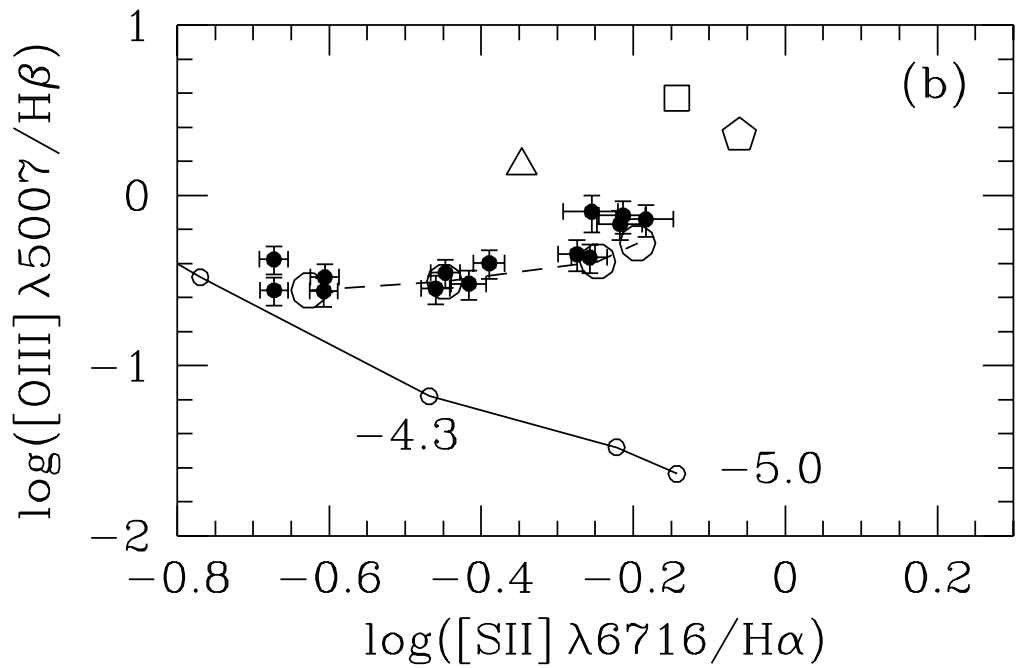
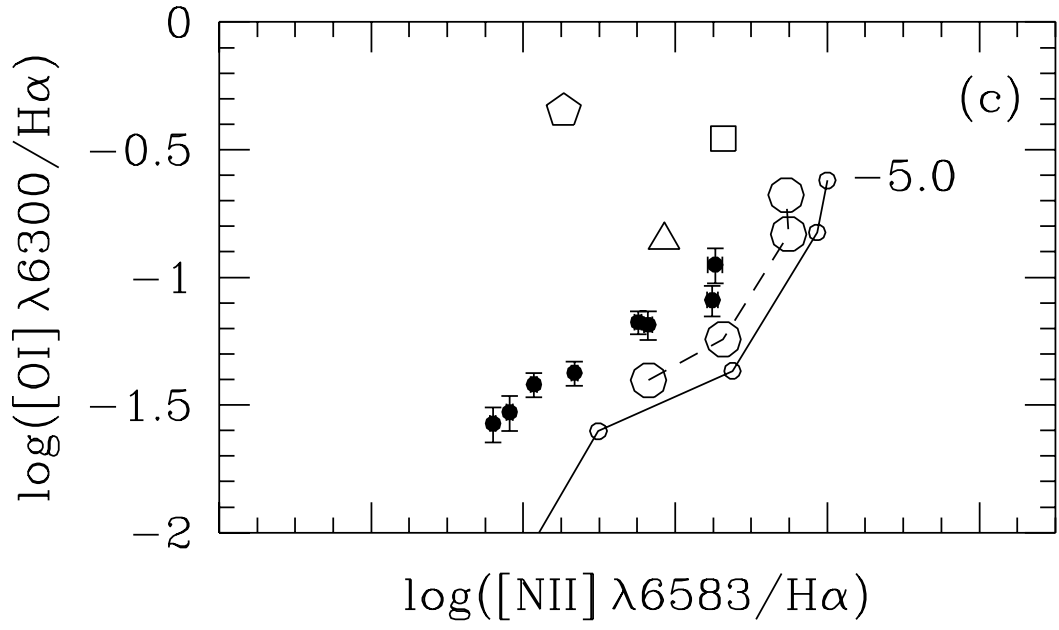
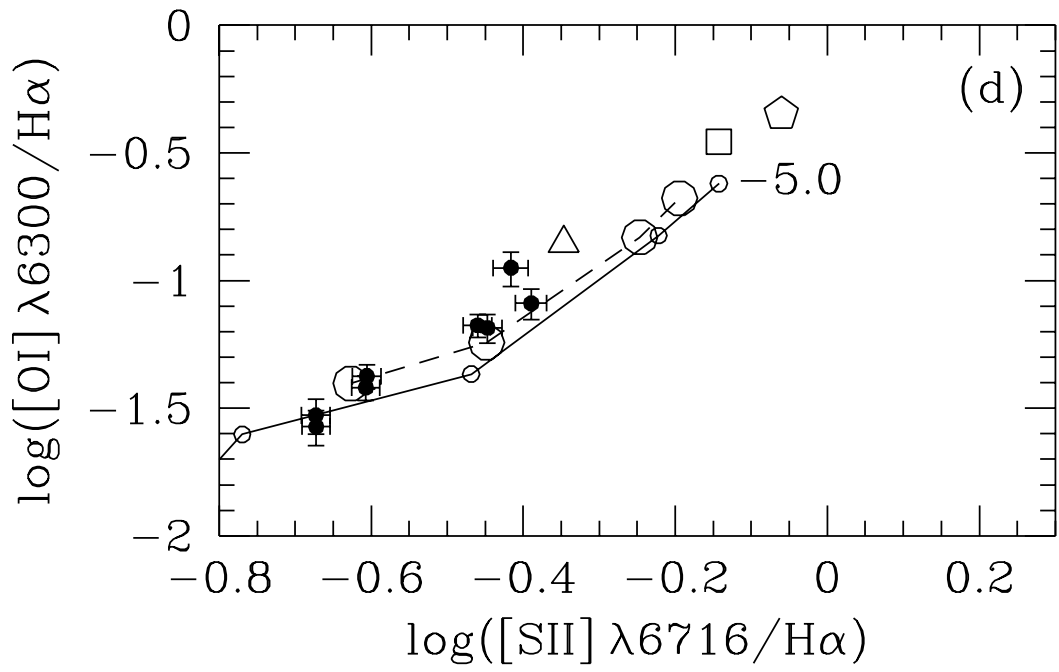
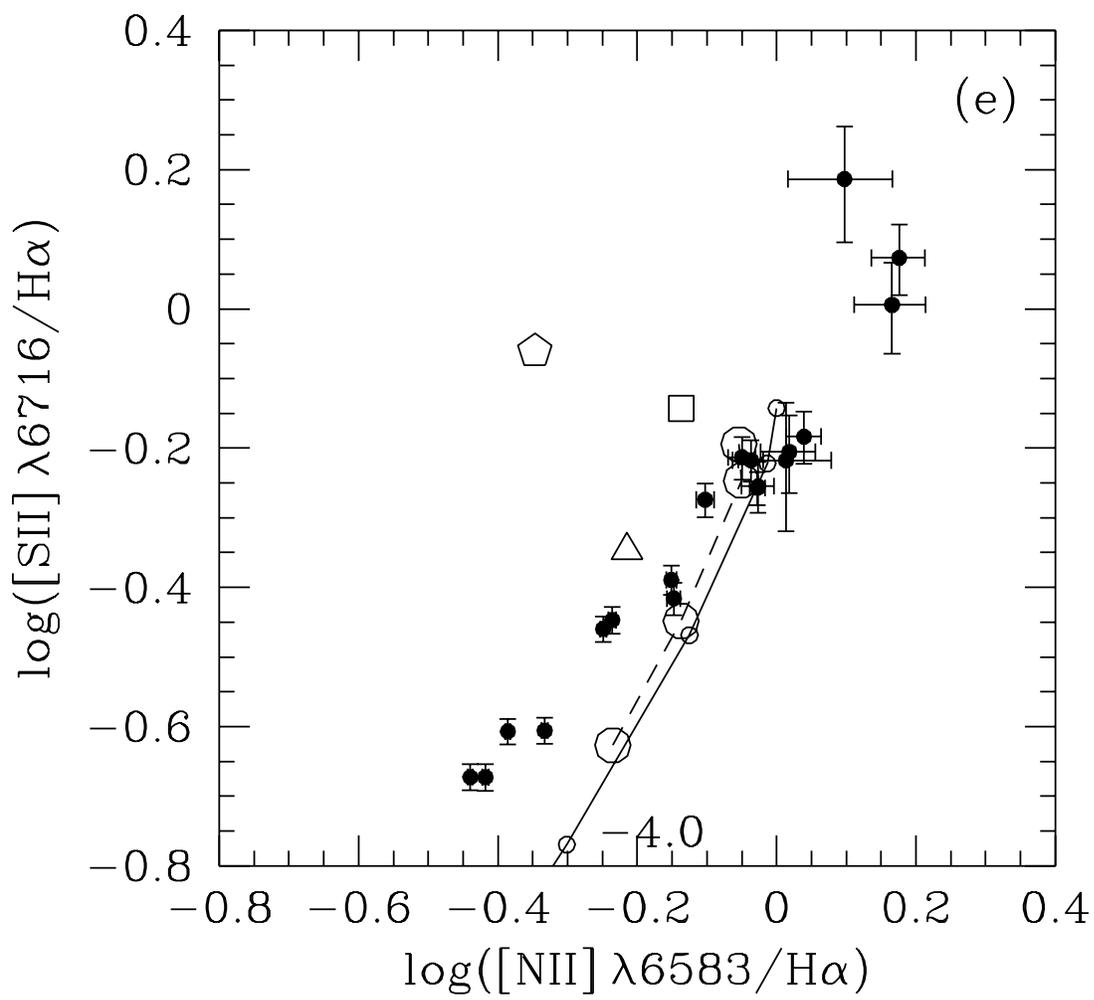
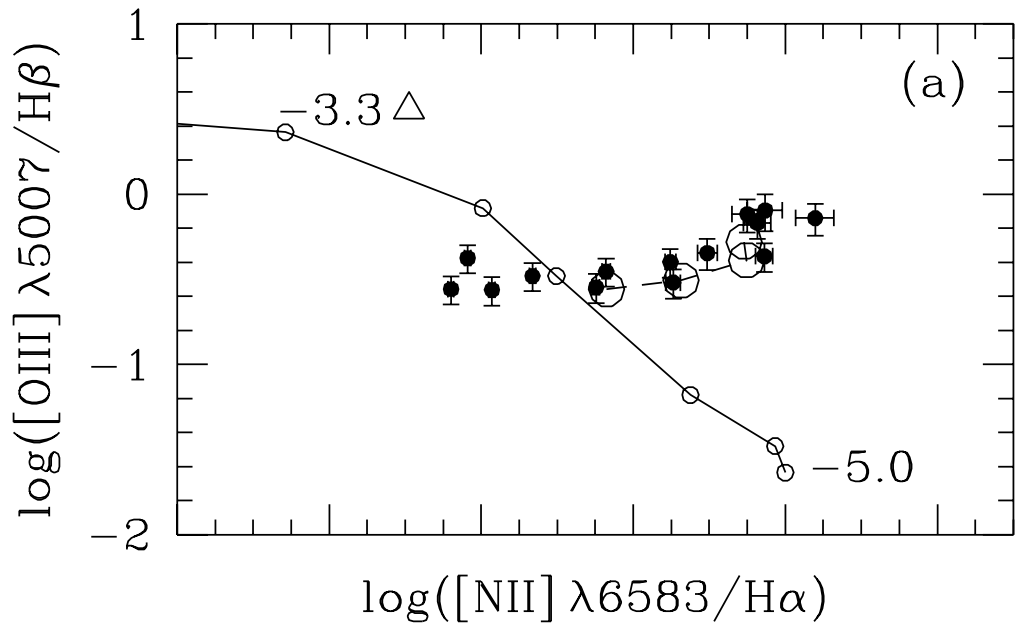
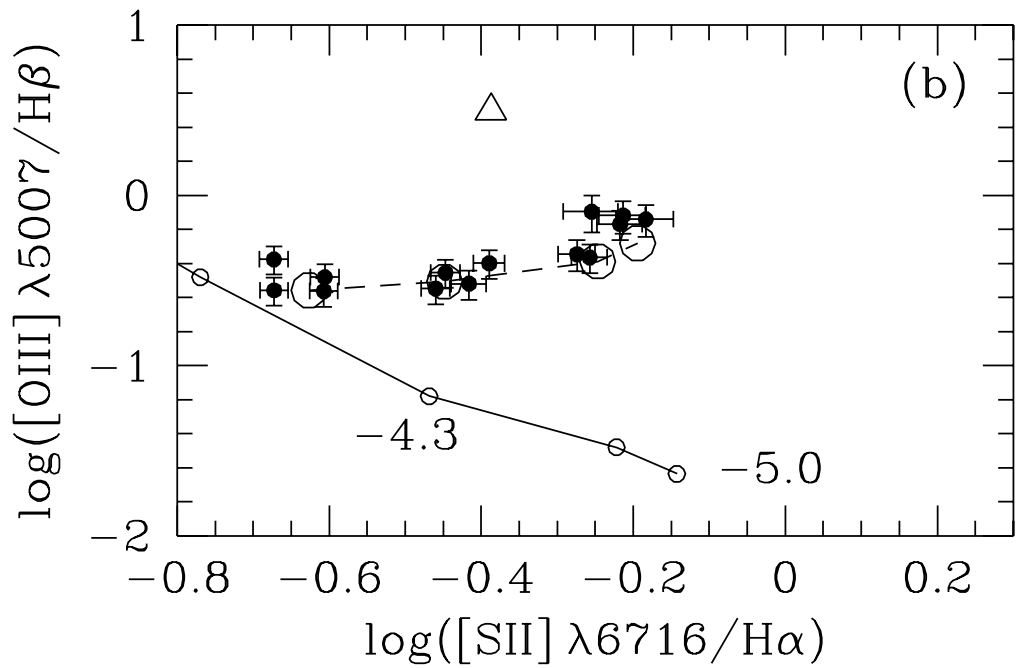
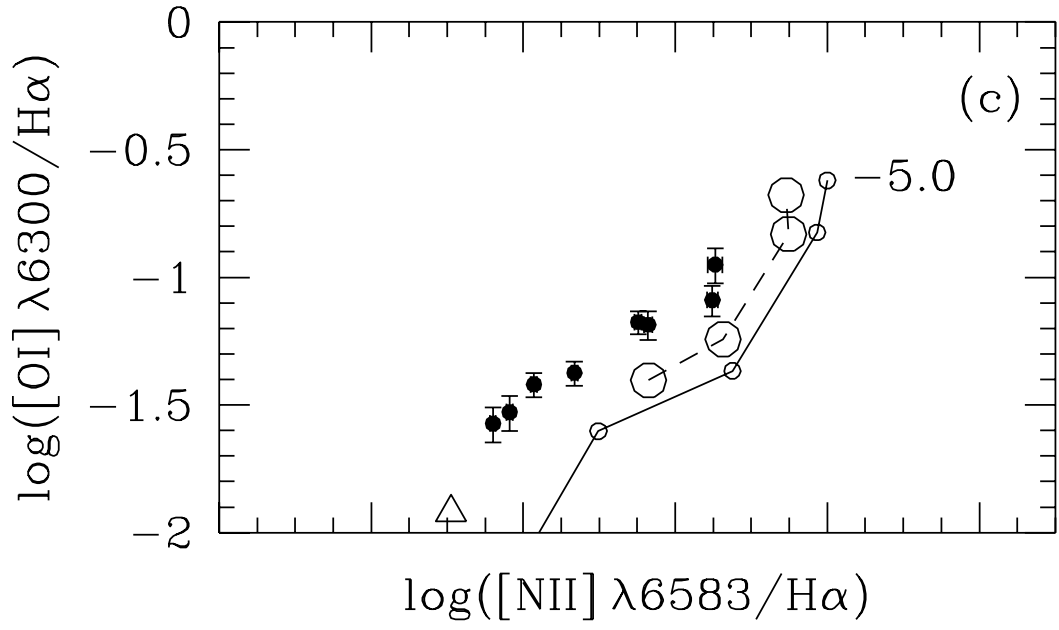
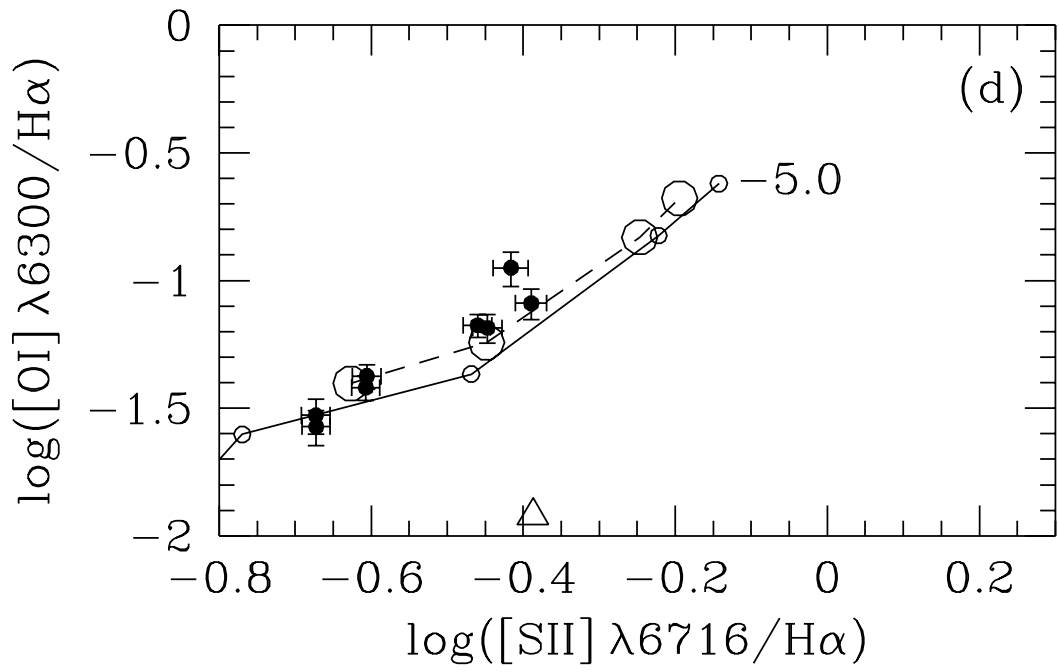
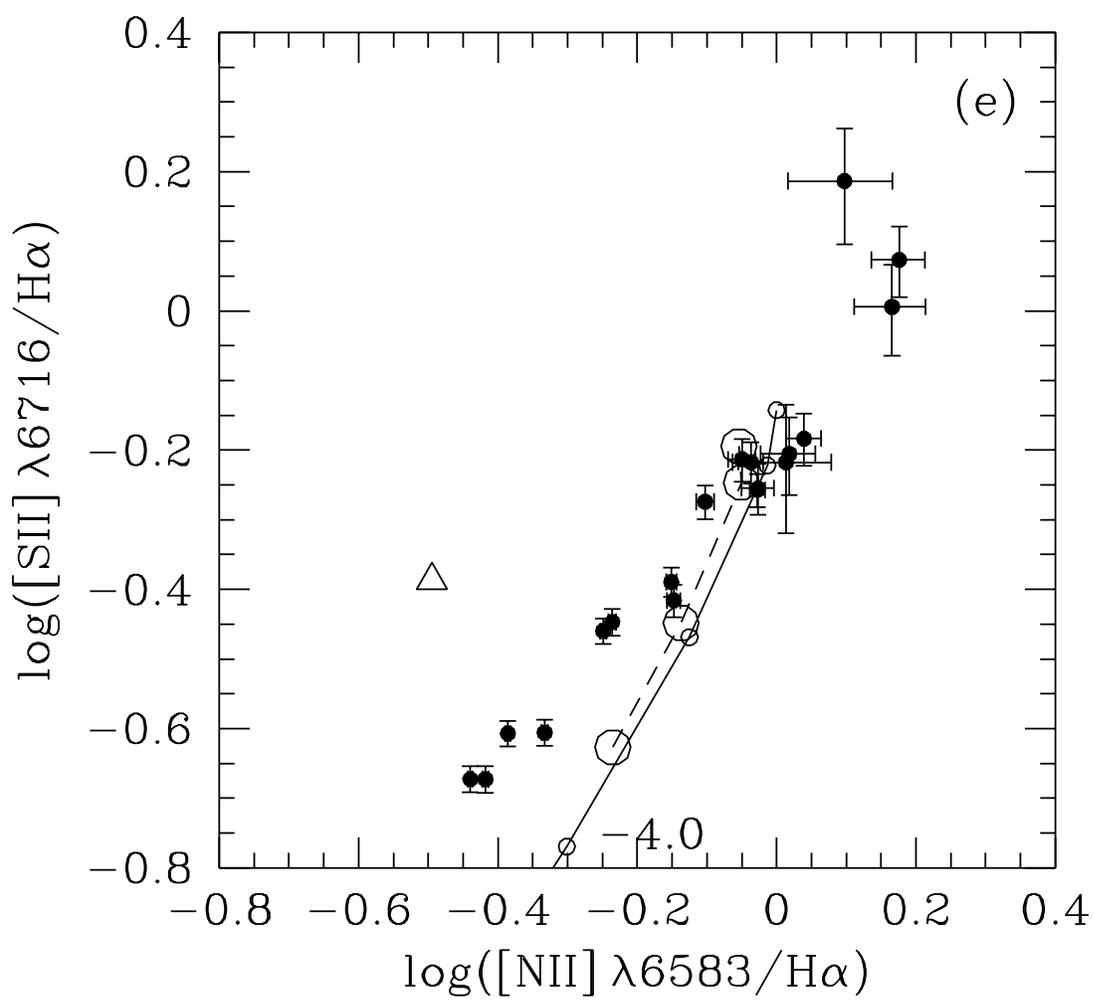
\end{document}